\documentclass[aps,prb,twocolumn,showpacs,10pt,floatfix,superscriptaddress,floatfix]{revtex4-1}
\usepackage[dvips]{graphics}
\usepackage{graphicx}
\usepackage[utf8]{inputenc}
\usepackage{amssymb}
\usepackage{epstopdf}
\usepackage{bm}
\usepackage{dcolumn}
\usepackage{epsfig}
\usepackage{latexsym}
\usepackage{amsmath,mathtools}
\usepackage{color}
\usepackage{array}
\usepackage{enumerate,dsfont}

\def\XXint#1#2#3{{\setbox0=\hbox{$#1{#2#3}{\int}$ }
\vcenter{\hbox{$#2#3$ }}\kern-.5\wd0}}

\newcommand{\va}{\bm a}
\newcommand{\ve}{\bm e}
\newcommand{\vn}{\bm n}
\renewcommand{\vr}{\bm r}
\newcommand{\vpsi}{\bm \psi}
\newcommand{\vK}{\bm K}
\newcommand{\vnu}{\bm \nu}
\newcommand{\vsigma}{\bm \sigma}
\newcommand{\vtau}{\bm \tau}

\makeatletter
\newcommand*{\rom}[1]{\expandafter\@slowromancap\romannumeral #1@}
\makeatother

\begin{document}

\title{Valley isospin of interface states in a graphene $pn$ junction in the quantum Hall regime}

\author{Luka Trifunovic}
\affiliation{Dahlem Center for Complex Quantum Systems and Physics Department, Freie Universit\"at Berlin, Arnimallee 14, 14195 Berlin, Germany}
\affiliation{Condensed Matter Theory Laboratory, RIKEN, Wako, Saitama 351-0198, Japan}

\author{Piet W.\ Brouwer}
\affiliation{Dahlem Center for Complex Quantum Systems and Physics Department, Freie Universit\"at Berlin, Arnimallee 14, 14195 Berlin, Germany}
\date{\today}

\begin{abstract}
In the presence of crossed electric and magnetic fields, a graphene ribbon has chiral states running along sample edges and along boundaries between $p$-doped and $n$-doped regions. We here consider the scattering of edge states into interface states, which takes place whereever the $pn$ interface crosses the sample boundary, as well as the reverse process. For a graphene ribbon with armchair boundaries, the evolution of edge states into interface states and {\em vice versa} is governed by the conservation of valley isospin. Although valley isospin is not conserved in simplified models of a ribbon with zigzag boundaries, we find that arguments based on isospin conservation can be applied to a more realistic modeling of the graphene ribbon, which takes account of the lifting of electron-hole degeneracy. The valley isospin of interface states is an important factor determining the conductance of a graphene $pn$ junction in a quantizing magnetic field.
\end{abstract}

\maketitle
\section{Introduction}

In the absence of scattering processes that cause a large momentum transfer, the low-energy electronic properties of graphene are described in terms of massless Dirac electrons near two inequivalent ``valleys'' at momenta $\vK$ and $\vK'$.\cite{novoselov2004,castroneto2009,dassarma2011} Since the corresponding four-component Dirac Hamiltonian is invariant with respect to ``rotations'' between the valleys, electrons can be assigned a {\em valley isospin}, which is described by a two-component spinor with quantization axis $\vnu$, such that $\vnu$ pointing in the positive (negative) $z$ direction corresponds to a fully valley-polarized state in the $\vK$ ($\vK'$) valley. In principle, the valley degree of freedom can be used to encode and process information, a concept that has given rise to the field of ``valleytronics''.\cite{schaibley2016} 

A measurement of the valley degree of freedom requires lifting of the valley degeneracy. Most notably, this occurs at the edges of a graphene sheet\cite{rycerz2007b} --- although other schemes, {\em e.g.}, involving the valley-dependent magnetic\cite{xiao2007} and optoelectronic effects,\cite{yao2008} can also be used. A robust valley dependence exists for the chiral edge states propagating along the boundaries of a graphene sheet in the lowest quantum Hall plateau, which have a well-defined valley isospin $\vnu$ if the lattice termination at the edge is regular.\cite{brey2006,beenakker2008} Consequentially, devices based on quantized-Hall graphene edge states have been proposed to detect and manipulate the valley degree of freedom in graphene, see, {\em e.g.}, Refs.\ \onlinecite{Rycerz2007,akhmerov2007} for two early examples. 

\begin{figure}
  \includegraphics[width=0.99\columnwidth]{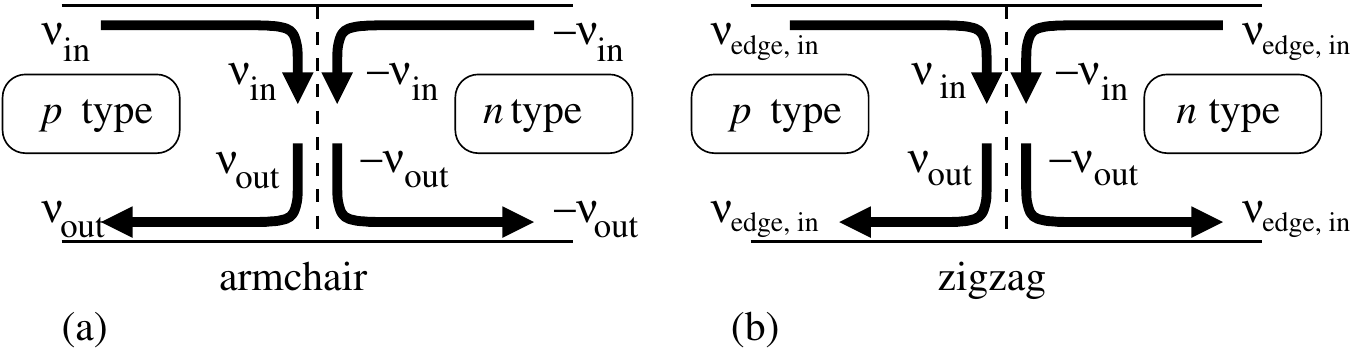}
  \caption{Schematic picture of a graphene $pn$ junction in the first quantized Hall plateau. (a) For an armchair nanoribbon chiral states coming in from the left and right (going out to the left and right) have opposite valley isospin $\vnu_{\rm in}$ and $-\vnu_{\rm in}$ ($\vnu_{\rm out}$ and $-\vnu_{\rm out}$), respectively. For sufficiently smooth potentials the valley isospin $\vnu$ is conserved and the conductance is determined by the overlap of valley isospins $\vnu_{\rm in}$ and $\vnu_{\rm out}$ of incoming and outcoming scattering states, see Eq.\ (\ref{eq:conductance}). (b) For a lattice model of a zigzag nanoribbon with nearest-neighbor hopping only, the incoming (outgoing) edge states on the left and right have the same isospin $\vnu_{\rm edge, in}$ ($\vnu_{\rm edge, out}'$). Valley isospin is not conserved in the transition between edge states and interface states. The conductance of the $pn$ junction is determined by the overlap of the isospins $\vnu_{\rm in}$ and $\vnu_{\rm out} $ of the interface states connected to incoming and outgoing scattering states in the same ($p$) side of the junction.}
\label{fig:1}
\end{figure}

A system that has received considerable theoretical\cite{abanin2007,tworzydlo2007,carmier2010,carmier2011,li2008,long2008,chen2011b,cavalcante2016,cohdnitz2016,fraessdorf2016,sekera2017,ma2018} and experimental attention is a graphene $pn$ junction in a quantizing magnetic field.\cite{williams2007,lohmann2009,ki2009,ki2010,woszczyna2011,williams2011,schmidt2013,matsuo2015b,klimov2015,matsuo2015,kumada2015,handschin2017,makk2018} The chiral edge states in such a $pn$ junction move in opposite directions in the $p$- and $n$-type regions. They feed into/flow out of two co-propagating valley-degenerate interface states at the $pn$ interface, see Fig.\ \ref{fig:1}. These interface states are also known as ``snake states'' because classical electron trajectories are curved such that they move alternatingly on the $p$- and $n$-sides of the $pn$ interface, a behavior reminiscent of the chiral states that propagate along zero-field contours in quantum Hall insulators in an inhomogeneous magnetic field.\cite{mueller1991,ye1995,reijniers2000,ghosh2008}

For a nanoribbon with armchair edges intervalley scattering is absent if the potential defining the $pn$ junction is smooth enough and the magnetic field is sufficiently weak (cyclotron radius much larger than lattice spacing). In that case the conductance of the $pn$ junction can be obtained from valley-isospin conservation arguments,\cite{tworzydlo2007}
\begin{equation}
  G = \frac{e^2}{h} (1 - \vnu_{\rm in} \cdot \vnu_{\rm out})
  \label{eq:conductance}
\end{equation}
where $\vnu_{\rm in} = \vnu_{p,\, {\rm in}} = - \vnu_{n,\, {\rm in}}$ and $\vnu_{\rm out} = \vnu_{p,\, {\rm out}} = -\vnu_{n,\, {\rm out}}$ are the valley isospin vectors for incoming and outgoing chiral edges states in the $p$ and $n$-type regions, respectively, respectively, see Fig.\ \ref{fig:1}a. On the other hand, for a $pn$ junction with zigzag edges, the two chiral edge states feeding into/coming out of the valley-degenerate interface state are reported to have the {\em same} valley isospin,\cite{tworzydlo2007,brey2006,goerbig2011} which rules out an isospin-conserving transition from edge state to interface state, see Fig.\ \ref{fig:1}b. This precludes the use of arguments invoking the conservation of valley isospin to determine the conductance of the $pn$ junction.\cite{tworzydlo2007} Nevertheless, the zigzag $pn$ junction {\em as a whole} was found to have well-defined valley-dependent transmission properties,\cite{tworzydlo2007} and it, too, has been proposed as a viable valleytronic device.\cite{sekera2017,handschin2017}

The absence of valley isospin conservation in these graphene $pn$ junctions is remarkable, because it exists no matter how smooth the scalar and vector potentials are. A possible origin of valley-isospin non-conserving scattering in graphene $pn$ junctions with zigzag edges was pointed out by Akhmerov {\em et al.},\cite{akhmerov2008} who analyzed a narrow zigzag $pn$ junction in zero magnetic field. Both without and with a magnetic field, a zigzag junction admits states localized near the sample edges.\cite{klein1994,fujita1996} As their momentum approaches the zone boundary, the transverse localization length of these states becomes of the order of the lattice constant. In the model studied by Akhmerov {\em et al.} --- a tight-binding model with nearest-neighbor hopping only---, the edge state momentum hits the zone boundary precisely at the $pn$ interface, {\em i.e.}, precisely where the Dirac point crosses the Fermi energy. The corresponding small spatial length scale can then supply the large momentum transfer required for valley-mixing scattering at the $pn$ interface.\cite{akhmerov2008}


An important point in this mechanism
is that the position at which edge states become maximally localized precisely coincides with the $pn$ interface. 
This indeed happens 
for the simple tight-binding model investigated in Ref.\ \onlinecite{akhmerov2008} and elsewhere,\cite{tworzydlo2007,handschin2017} in which graphene is described as a hexagonal lattice with nearest-neighbor hopping. Such models have a sublattice antisymmetry, which pins the energy of the maximally localized edge state to the Dirac point. In more realistic models, the sublattice antisymmetry is broken, {\em e.g.}, by next-nearest-neighbor hopping or by on-site potentials at the sample edges. In that case there can be a finite energy difference $\delta U$ between the energy of the zone-boundary-localized state and the Dirac point.\cite{sasaki2006} As a result, the chiral edge states of the lowest quantized Hall plateau are not localized on an atomic length scale in the vicinity of the $pn$ interface and there is no short length scale that can facilitate intervalley scattering. Instead, intervalley scattering (if any) takes place well away from the $pn$ interface and the two chiral edge states incident on the $pn$ interface have {\em opposite} valley isospin.


In this article we consider the valley isospin of interface states in a graphene $pn$ junction in the first quantized Hall plateau with zigzag edges, comparing models with and without sublattice antisymmetry. For models with sublattice antisymmetry, using a combination of numerical and analytical arguments we calculate the isospin $\vnu_{\rm in} = \vnu_{p,\, {\rm in}} = - \vnu_{n,\, {\rm in}}$ ($\vnu_{\rm out} = \vnu_{p,\, {\rm out}} = - \vnu_{n,\, {\rm out}}$) of the interface states, where the indices $p$ and $n$ refer to interface states that evolves out of (into) the chiral edge states at the $p$ and $n$ side of the junction, respectively, see Fig.\ \ref{fig:1}b. Once the valley isospin of the interface states is known, Eq.\ (\ref{eq:conductance}) is still applicable, provided the isospins $\vnu_{\rm in}$ and $\vnu_{\rm out}$ are taken to be the isospins of the interface states (as defined above), not of the chiral edge states. Comparing our results for the conductance of a $pn$ junction with zigzag edges with those obtained at zero magnetic field,\cite{wakabayashi2002,akhmerov2008} we find no magnetic field dependence of the conductance, despite the vastly different limits involved (metallic ribbon with a finite-size gap vs.\ quantized Hall insulator). Whereas we do not have a formal calculation to prove this observation, we attribute it to the understanding that in the presence of sublattice antisymmetry the intervalley scattering is essentially a short-distance effect taking place within a few lattice spacings from where the $pn$ interface meets the sample edge,\cite{akhmerov2008} whereas the magnetic field affects electrons on a much longer length scale.

For the (more realistic) models without sublattice antisymmetry, we observe that, 
in contrast to what was found in the absence of sublattice antisymmetry,
chiral edge states impinging on the $pn$ interface sublattice have {\em opposite} isospin $\vnu_{\rm in} = \vnu_{p,\, {\rm in}} = -\vnu_{n,\, {\rm in}}$ {\em even for a zigzag edge}. 
The same applies for outgoing states. For a sufficiently smooth $pn$ junction this means that Eq.\ (\ref{eq:conductance}) can still be used to describe the conductance of a $pn$ junction, without having to redefine the meaning of $\vnu_{\rm in}$ and $\vnu_{\rm out}$. For models without sublattice antisymmetry the conductance is qualitatively different from what is obtained based on the simple tight-binding model with nearest-neighbor hopping only.\cite{tworzydlo2007,handschin2017} In particular, we find that zigzag nanoribbons of even and odd width have the same conductance $G = 2 e^2/h$, in contrast to Refs.\ \onlinecite{tworzydlo2007,handschin2017}, who find that $G = 2e^2/h$ for ribbons with even width and $G = 0$ for ribbons with odd width. 

The importance of sublattice-symmetry-breaking terms can be estimated by considering the role of a next-nearest-neighbor hopping amplitude $t'$. In the absence of potentials at the sample edges, one has $\delta U = t'$;\cite{sasaki2006} localized potentials at the sample edge can further change this shift. Experimentally $t'$ is estimated at $t' \approx 0.3$ eV.\cite{kretinin2013} Therefore, a nonzero value of $\delta U$ is a relevant perturbation for $pn$ junctions if the applied potential remains smaller than $\delta U$ in a region (much) larger than the lattice constant $a$ around the $pn$ interface. This is the case if the over-all potential drop across the $pn$ junction is well below $\delta U$ (``small-amplitude junction''), or if the in-plane electric field $|E| \ll \delta U/e$ (distance being measured in units of the lattice spacing). Both conditions should realistically be met in most experiments.

For the model with sublattice symmetry, for which the valley isospins $\vnu_{\rm in}$ and $\vnu_{\rm out}$ of the interface states do not follow from isospin conservation arguments, we numerically obtain $\vnu_{\rm in}$ and $\vnu_{\rm out}$ of the interface states by solving for scattering states in a stub geometry, for which the $pn$ junction and the interface states extend into the lead. This allows our calculation to go beyond numerical and analytical studies of the conductance of zigzag $pn$ junctions as a whole,\cite{wakabayashi2002,akhmerov2008,handschin2017} in which no information specific to interface states could be obtained. It also allows us to apply our results to graphene $pn$ junctions with non-parallel edges.

Our results also explain the remarkably strong parameter dependence of the conductance of a disordered $pn$ junction with zigzag edges that has been observed in previous numerical studies on models with nearest-neighbor hopping only. In particular, Ref.\ \onlinecite{tworzydlo2007} found that the conductance of a $pn$ junction with zigzag edges depended very strongly on the value of the Fermi energy, whereas there was no such dependence in the case of a $pn$ junction with armchair edges. Reference \onlinecite{handschin2017} found a strong dependence of the conductance of a disordered zigzag $pn$ junction on the precise position of the $pn$ interface. We can understand these results by noting that the isospins $\vnu_{\rm in}$ and $\vnu_{\rm out}$ of the interface state depend sensitively on the precise position of the $pn$ interface: Already a shift of the intersection of the $pn$ interface and the sample edge by a distance of the order of a lattice constant rotates $\vnu_{\rm in}$ or $\vnu_{\rm out}$ by a large angle. A (smooth) disorder potential effectively causes a random shift of the positions of the intersections of the $pn$ interface and the two sample edges, corresponding to random rotations of the valley isospin $\vnu_{p,\, {\rm in}}$ and $\vnu_{p,\, {\rm out}}$. A similar strong dependence on the position of the $pn$ interface was observed by Akhmerov {\em et al.}\ for zigzag $pn$ junctions in the absence of a magnetic field.\cite{akhmerov2008}

The article is organized as follows: In Sec.~\ref{sec:model} we review the concept of valley isospin and derive constraints for the isospins $\vnu_{\rm in}$ and $\vnu_{\rm out}$ of the interface states for high-symmetry positions of the $pn$ interface and for a simplified tight-binding model of graphene with nearest-neighbor hopping only. In Sec.\ \ref{sec:numerics} we show numerical results for the isospin of interface states. Section \ref{sec:extensions} describes applications of our theory to the conductance of $pn$ junctions with various boundary terminations and compares with the zero-energy theory of Ref.\ \onlinecite{akhmerov2008}.
We conclude in Sec.\ \ref{sec:conclusion}.

\section{Valley isospin and symmetry analysis}\label{sec:model}

\subsection{Valley isospin}

\begin{figure}
  \includegraphics[width=0.7\columnwidth]{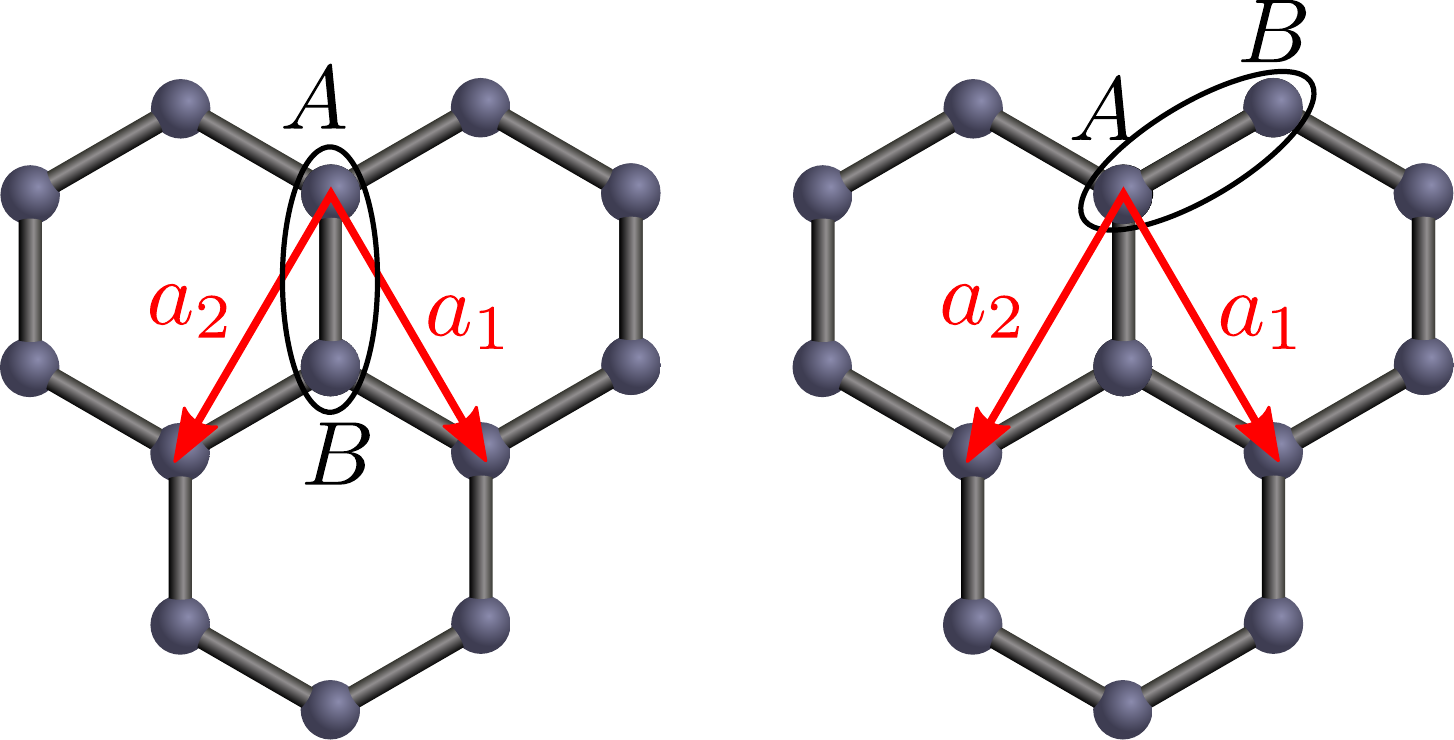}
  \caption{Hexagonal lattice with its two-atom unit cell. The primitive lattice vectors are denoted $\va_1$ and $\va_2$. The right panel shows a choice of the unit cell that is rotated anti-clockwise by an angle $2 \pi/3$. }
\label{fig:lattice}
\end{figure}

Graphene has a hexagonal arrangement of Carbon atoms, as shown in Fig.\ \ref{fig:lattice}. It can be described as a triangular lattice with a two-atom basis, labeled ``$A$'' and ``$B$''. The low-energy physics of conduction electrons in graphene takes place for momenta near one of two inequivalent corners of the Brillouin zone. The momenta of these corners or ``valleys'' are labeled $\vK$ and $\vK'$. We choose the primitive lattice vectors $\va_1$ and $\va_2$ such, that they obey\cite{castroneto2009}
\begin{equation}
  e^{i \vK \cdot \va_1} = e^{-i \vK \cdot \va_2} = e^{-i \vK' \cdot \va_1} =
  e^{i \vK' \cdot \va_2} = e^{-2 \pi i/3}.
\end{equation}
We neglect spin-orbit coupling, which is very weak in graphene, and do not consider the electron spin explicitly, except for the over-all factor two in Eq.\ (\ref{eq:conductance}). Electronic states are represented by a two-component pseudospinor $\vpsi(\vr)$, where the pseudospinor degree of freedom corresponds to the sublattice structure and $\vr$ is a lattice vector.

The pseudospinor of a low-energy state can be written as a sum of contributions at the two valleys,
\begin{align}
  \vpsi(\vr) =&\, \begin{pmatrix} \psi_{A}(\vr) \\ \psi_{B}(\vr) \end{pmatrix}
  \nonumber \\ =&\,
  \begin{pmatrix} \phi_{A}(\vr) \\ \phi_{B}(\vr) \end{pmatrix}
  e^{i \vK \cdot \vr} +
  \begin{pmatrix} \phi_{A}'(\vr) \\ \phi_{B}'(\vr) \end{pmatrix}
  e^{i \vK' \cdot \vr},
\end{align}
where $\phi_{A,B}(\vr)$ and $\phi_{A,B}'(\vr)$ are slow functions of the lattice vector $\vr$. A low-energy state $\vpsi(\vr)$ has a valley isospin $\vnu$ if the corresponding four-component spinor $\Psi$, defined by
\begin{equation}
  \Psi(\vr) = \begin{pmatrix}
  \phi_{A}(\vr) \\ \phi_{B}(\vr) \\ -\phi_{B}'(\vr) \\ \phi_{A}'(\vr) \end{pmatrix},
\end{equation}
obeys the condition\cite{beenakker2008}
\begin{equation}
  (\vnu \cdot \vtau) \Psi(\vr) = \Psi(\vr)
\end{equation}
for all $\vr$, where the $\tau_{j}$, $j=1,2,3$, are Pauli matrices acting on the valley degree of freedom. The two-component pseudospinor $\vpsi(\vr)$ of a such state with well-defined valley-isospin $\vnu$ has the form
\begin{align}
  \vpsi(\vr) =&\, \begin{pmatrix} \psi_{A}(\vr) \\ \psi_{B}(\vr) \end{pmatrix}
  \nonumber \\ =&\,
  \nu_1 \begin{pmatrix} \phi_A(\vr) \\ \phi_B(\vr) \end{pmatrix} e^{i \vK \cdot \vr}
  + \nu_2 \begin{pmatrix} \phi_B(\vr) \\ -\phi_A(\vr) \end{pmatrix} e^{i \vK' \cdot \vr},
  \label{eq:phases}
\end{align}
where the two amplitudes $\nu_1$ and $\nu_2$ form the two-component valley spinor corresponding to the isospin vector $\vnu$. (Note that the two-component spinor $(\nu_1,\nu_2)$ is defined up to a phase factor only.)

It is important to point out that the valley isospin $\vnu$ depends on the choice of the origin and of the unit cell. A translation of the origin from a lattice position $O$ to another lattice position $\bar O$ rotates the valley isospin by the angle $\theta_{O\bar O} = \vK \cdot \vr_{O\bar O}$ around the $z$ axis,
\begin{equation}
  \vnu \to \bar \vnu = {\cal R}_{z,\theta_{O \bar O}} \vnu,
  \label{eq:translation}
\end{equation}
where $\vr_{O\bar O}$ is the displacement vector pointing from $O$ to $\bar O$.
[The rotation angle can be calculated with the help of Eq.\ (\ref{eq:phases}).] Similarly, an anticlockwise rotation of the unit cell by an angle $2 \pi/3$, see Fig.\ \ref{fig:lattice}, rotates $\vnu$ by an angle $2 \pi/3$ around the $z$ axis, with a 
with a simultaneous change $(\phi_A,\phi_B) \to (\bar \phi_A,\bar \phi_B) = (\phi_A e^{-2 \pi i/3}, \phi_B)$ of the sublattice pseudospinor.

Within the Dirac equation description, valley isospin is a constant of the motion, since the Dirac Hamiltonian commutes with the valley isospin operator $\vtau$. Isospin rotation symmetry is broken at the sample edges. In general a boundary without time-reversal symmetry-breaking perturbations has a boundary condition of the form\cite{akhmerov2007,akhmerov2008b}
\begin{equation}
  \Psi = (\vnu_{\rm b} \cdot \vtau)(\vn_{\rm b} \cdot \vsigma) \Psi,
  \label{eq:generalboundary}
\end{equation}
where the $\sigma_j$, $j=1,2,3$, are Pauli matrices acting on the sublattice (pseudospin) degree of freedom and  $\vnu_{\rm b}$ and $\vn_{\rm b}$ are two unit vectors characteristic of the boundary termination. The vector $\vn_{\rm b}$ must be perpendicular to the boundary normal; no a priori constraints apply to the vector $\vnu_{\rm b}$.\cite{akhmerov2008b} 

\begin{figure}
  \includegraphics[width=0.9\columnwidth]{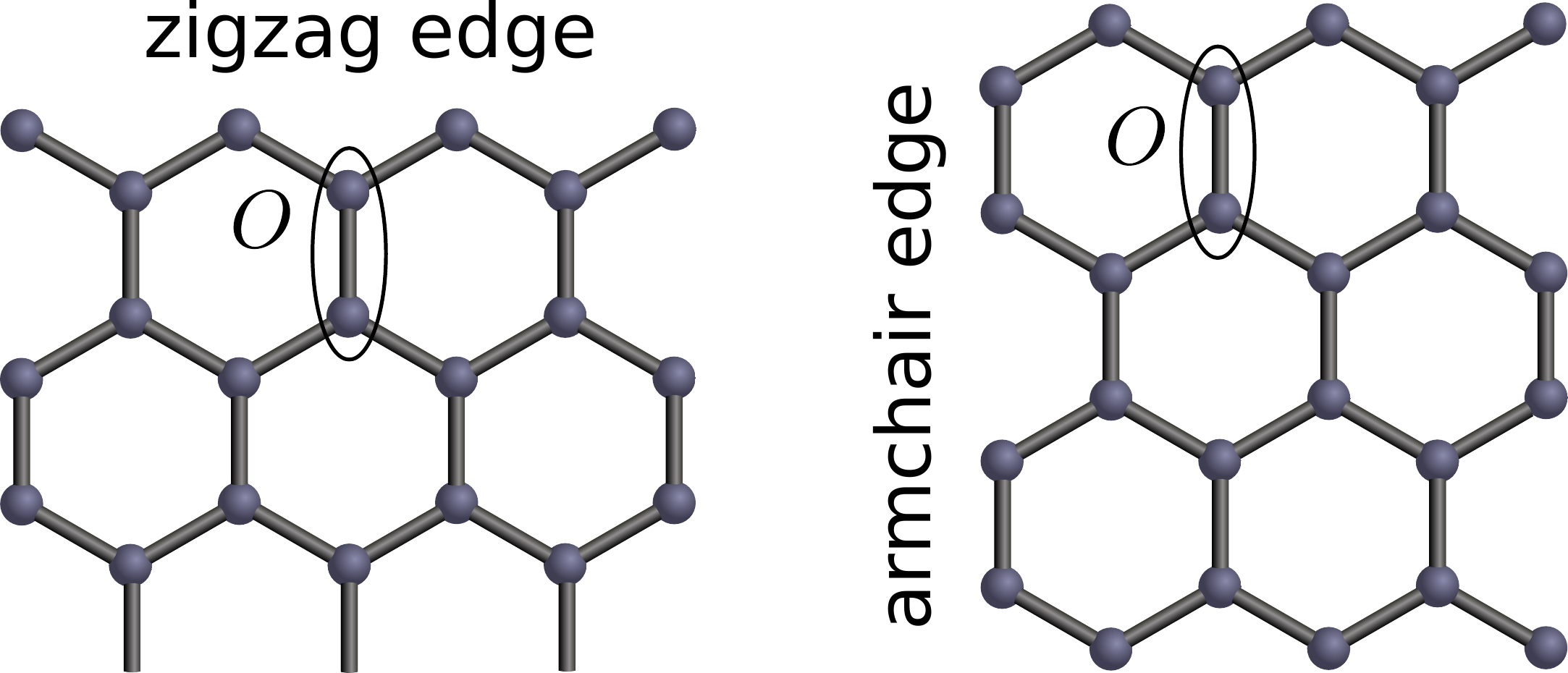}
  \caption{Choice of the origin $O$ and orientation of the two-atom unit cell corresponding to the boundary conditions (\ref{eq:zigzagboundary}) and (\ref{eq:armchairboundary}) for zigzag and armchair boundary conditions.}
\label{fig:boundary}
\end{figure}

An explicit form for the boundary conditions can be obtained in a tight-binding description with nearest-neighbor hopping only. For such a model one finds that at a zigzag edge one has the boundary condition\cite{akhmerov2008b} 
\begin{equation}
  \Psi = \pm \tau_z \sigma_z \Psi, \label{eq:zigzagboundary}
\end{equation}
if the two-atom unit cell is oriented perpendicular to the boundary, see Fig.\ \ref{fig:boundary}. The $+$ ($-$) sign applies to a zigzag boundary for which the outermost atoms are on the $A$ ($B$) sublattice. Equation (\ref{eq:zigzagboundary}) corresponds to the general case (\ref{eq:generalboundary}) with $\vnu_{\rm b} = \pm \vn_{\rm b} = \ve_z$. In the nearest-neighbor model the boundary condition for an armchair termination reads
\begin{equation}
  \Psi = \tau_y \sigma_y \Psi, \label{eq:armchairboundary}
\end{equation}
if the two-atom unit cell is oriented parellel to the boundary and the origin $O$ is chosen at the interior boundary of the outermost hexagon, see Fig.\ \ref{fig:boundary}.

\subsection{$pn$ junctions at the first quantized Hall plateau}

A graphene $pn$ junction at the first quantized Hall plateau has non-degenerate
chiral modes running along the sample edges, as shown schematically in Fig.\
\ref{fig:1}, as well as two valley-degenerate co-propagating chiral interface
modes. The edge modes have a well-defined valley isospin
$\vnu$.\cite{tworzydlo2007} The boundary conditions (\ref{eq:zigzagboundary})
and (\ref{eq:armchairboundary}) for the nearest-neighbor hopping model fix the
direction of the valley isospin $\vnu$ to be parallel to the $z$ or $y$ axis
for zigzag or armchair termination, but do not specify the direction of $\vnu$.
For the symmetry arguments that follow below it is not necessary to know the
sign of $\vnu$. However, for definiteness we will make use of the result of
numerical calculations of a lattice model with nearest-neighbor hopping, which
will be discussed in more detail in the next Section. For such a model one
finds that $\vnu = -\ve_z$ ($\ve_z$) for a chiral zigzag edge with outermost
atoms of $A$ ($B$) type, where $\ve_z$ is the unit vector in the $z$ direction.
For a junction with armchair termination one finds $\vnu = -\ve_y$ ($\ve_y$)
for an edge state moving in the direction $A \to B$ ($B \to A$) with respect to
a two-atom unit cell oriented parallel to the edge. If sublattice antisymmetry is broken, the valley isospin of a chiral mode at a zigzag edge changes sign in either the $p$-type region or the $n$-type region, depending on the sign of the sublattice-antisymmetry-breaking perturbation $\delta U$, see the discussion in Sec.\ \ref{sec:numerics}.

\subsection{Symmetry analysis: general case}

Whereas the valley isospin of the chiral edge states is determined by the boundary conditions, the valley isospin of the interface states has to follow from different considerations. If the $pn$ interface is placed symmetrically with respect to the sample boundary, the possible values of the valley isospin for the interface states can be constrained by symmetry arguments. 

\begin{figure}
  \includegraphics[width=\columnwidth]{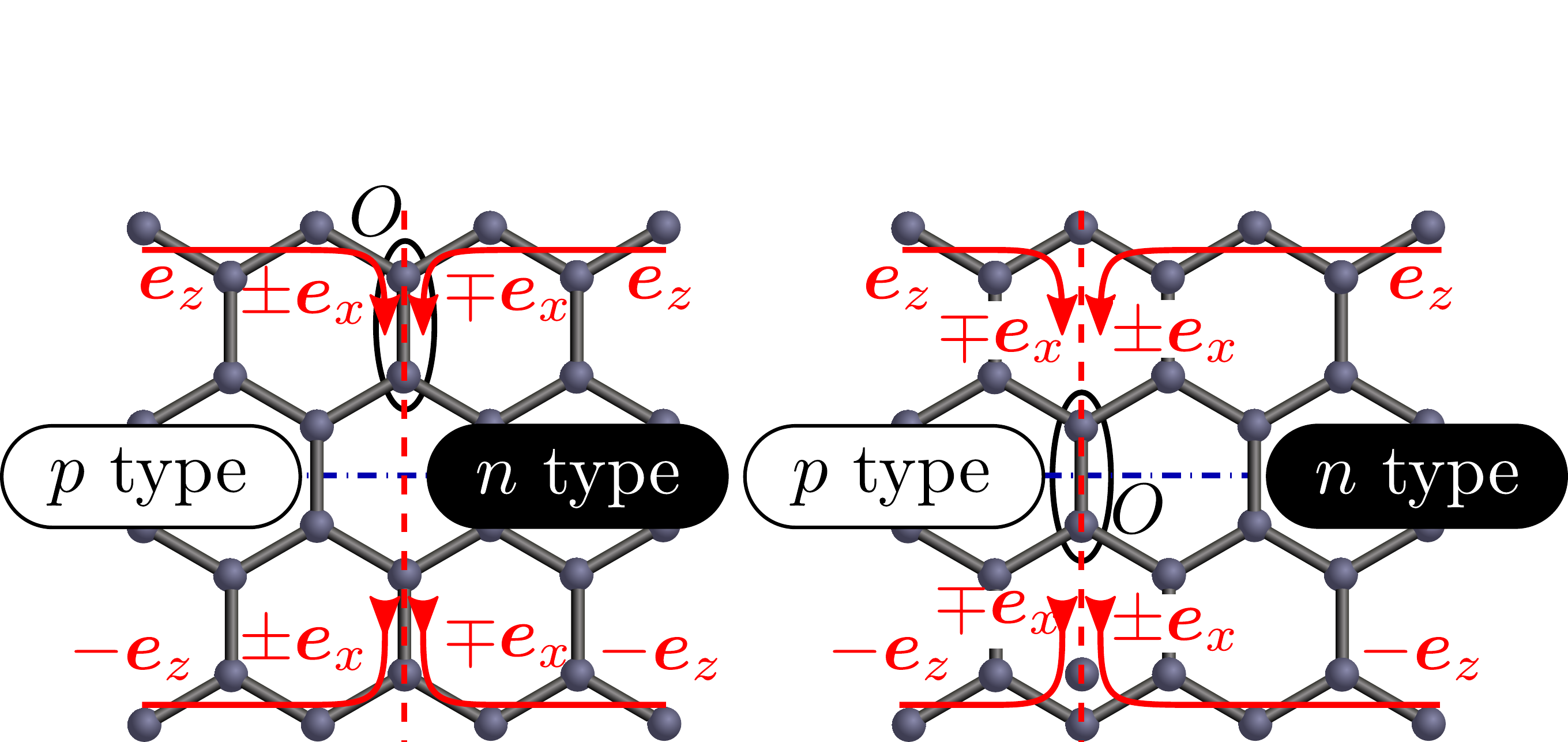}
  \caption{Zigzag $pn$ junctions with two high-symmetry positions of the $pn$ interfaces (dashed) and with mirror axis parallel to edge (dot-dash). The unit cell labeled $O$ is compatible with the boundary condition (\ref{eq:zigzagboundary}) at both boundaries. The unit cells are positioned symmetrically with respect to mirror inversion in the $pn$ interface. Valley isospin directions of edge states and interface states are indicated for a nearest-neighbor tight-binding model. The isospin directions of the interface states are fixed by symmetry arguments, see main text, up to an over-all sign, which is determined by numerical simulations (see Sec.\ \ref{sec:numerics}).}
\label{fig:junction_mirror_zigzag}
\end{figure}

{\em Junction with mirror axis parallel to sample edge.---} A symmetry axis parallel to the sample edge allows one to relate valley isospins for incoming and outgoing scattering states. This situation is shown schematically in Figs.\ \ref{fig:junction_mirror_zigzag} and \ref{fig:junction_mirror_armchair} for $pn$ junctions with zigzag and with armchair edges, respectively. Although the magnetic field breaks the mirror symmetry, the system remains invariant under the combined operation ${\cal MT}$ of a mirror reflection and time-reversal. This symmetry operation corresponds to the valley isospin change
\begin{equation}
  \vnu \to \bar \vnu = \left\{ \begin{array}{ll} {\cal M}_z \vnu & \mbox{zigzag}, \\
  {\cal M}_y \vnu & \mbox{armchair}, \end{array} \right.
  \label{eq:mirror_parallel}
\end{equation}
where ${\cal M}_y$ and ${\cal M}_z$ denote mirror reflection in the $y=0$ and $z=0$ planes in the Bloch sphere, respectively. The valley isospin $\bar \vnu$ is calculated with respect to the mirror image $\bar O$ of the origin $O$, see Figs.\ \ref{fig:junction_mirror_zigzag} and \ref{fig:junction_mirror_armchair}. Note that the prescription $\vnu \to {\cal M}_y \vnu$ is consistent with the observation of the previous subsection that the valley isospin of a chiral state at an armchair edge state points along the $y$ axis for the simple nearest-neighbor model and changes sign upon changing the direction of propagation. The prescription $\vnu \to {\cal M}_z \vnu$ is consistent with the observation that the valley isospin of a chiral state at a zigzag edge points along the $z$ axis and changes sign if the outermost atoms change from the $A$ sublattice to the $B$ sublattice.

{\em Conductance.---} The above symmetry argument is sufficient to determine the conductance of a $pn$ junction with armchair edges and a mirror axis parallel to the edge.\cite{tworzydlo2007} One first considers a ``minimal'' junction with a width alternating between one and two hexagons --- which we refer to as a width of four ``half-hexagons'' ---, for which a symmetric position of the origin $O = \bar O$ compatible with the boundary condition (\ref{eq:armchairboundary}) on both edges. Since isospin is conserved at the $pn$ interface for armchair boundaries, the valley isospin of the interface state coming from the $p$-type side of the junction has valley isospin $-\ve_y$, which is the same as the isospin of the interface state that evolves into a chiral edge state moving out to $n$-type side of the junction. We conclude that this $pn$ junction is perfectly transmitting, $G=2 e^2/h$. The conductance of armchair $pn$ junctions of arbitrary width can then be found using the fact that valley isospin rotates by an angle $\theta_{O \bar O} = \vK \cdot \vr_{O \bar O}$ around the $z$ axis if the origin is translated from $O$ to $\bar O$. The rotation corresponding to a width increase of half a hexagon is of magnitude $2 \pi/3$, from which it follows that $G = 2 e^2/h$ for all armchair junctions that can be obtained from the minimal junction by adding $3 n$ rows of half hexagons with $n$ integer, and $G = (e^2/h) (1 - \cos \pi/3) = e^2/2h$ otherwise.\cite{tworzydlo2007}

\subsection{Symmetry analysis: Models with sublattice antisymmetry}

The presence of other spatial symmetries, such as a mirror axis perpendicular to the edge or an inversion center, does not constrain the valley isospin of chiral edge or interface states, because such symmetries are broken by the simultaneous application of electric and magnetic fields at the $pn$ junction. These additional symmetries can be combined however, with a sublattice antisymmetry ${\cal C}$, 
\begin{equation}
  H = - \sigma_3 H \sigma_3,
\end{equation}
which is an exact antisymmetry for the simple nearest-neighbor model of graphene, but not for more realistic models. Here $\sigma_3$ is a Pauli matrix acting on the sublattice (pseudospinor) degree of freedom. The combination of the sublattice antisymmetry and a mirror symmetry with a mirror axis parallel to the $pn$ interface or with an inversion center leads to strong constraints on the valley isospin of the interface states and on the over-all conductance of the $pn$ junction at half filling ({\em i.e.}, at zero energy), as we now discuss.

\begin{figure}
  \includegraphics[width=0.9\columnwidth]{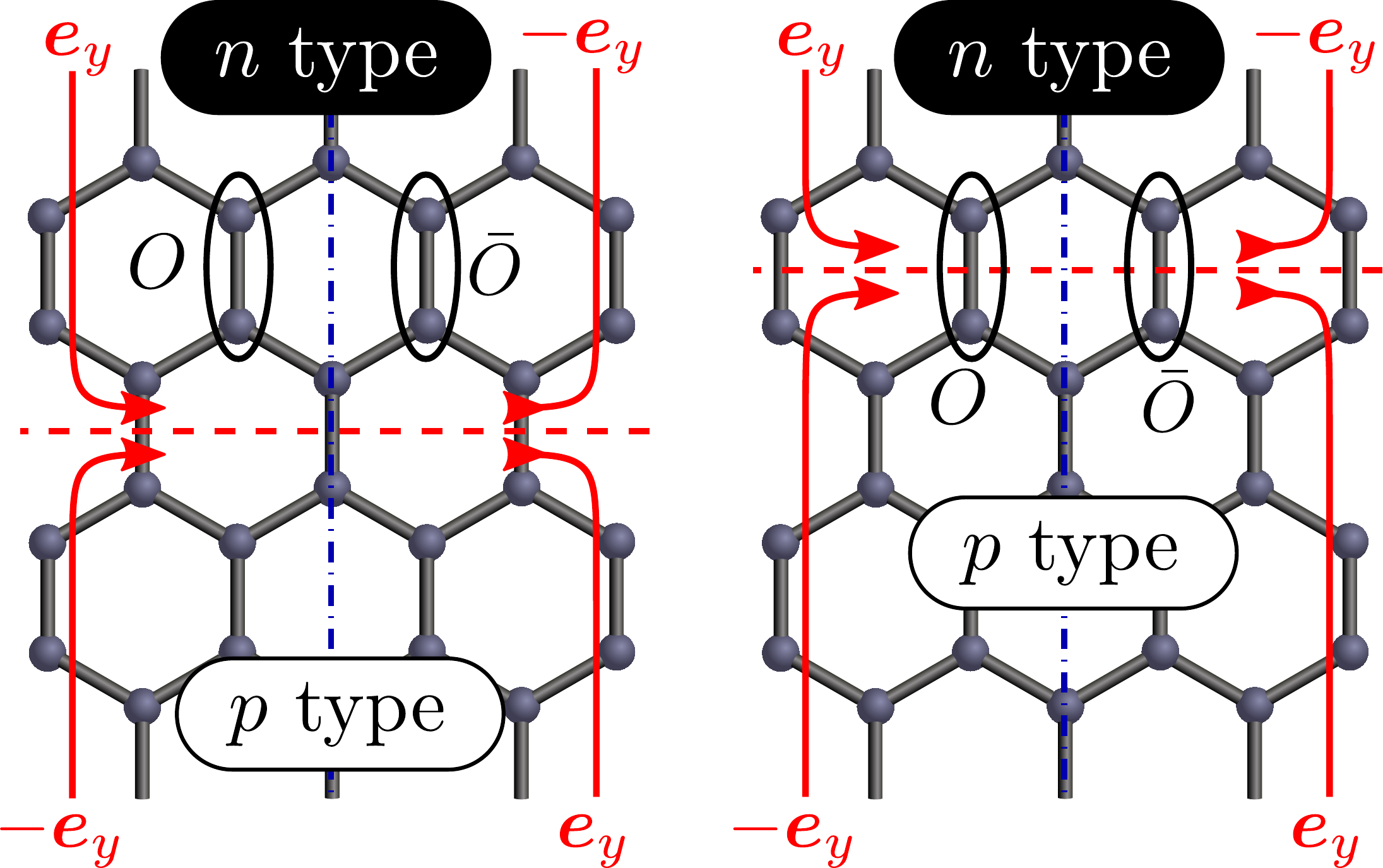}
  \caption{Armchair $pn$ junctions with high-symmetry positions of the $pn$ interfaces (dashed) and with mirror axis parallel to edge (dot-dash). Valley isospin labels are for a nearest-neighbor tight-binding model. They are given with respect to the origins $O$ and $\bar O$ for incoming and outgoing scattering states, respectively. (The origins $O$ and $\bar O$ are interchanged under mirror reflection in the mirror axis parallel to the edge.) Isospin directions are fixed by symmetry arguments, see main text, up to an over-all sign, which is determined by numerical simulations (see Sec.\ \ref{sec:numerics}).}
\label{fig:junction_mirror_armchair}
\end{figure}

{\em Mirror axis perpendicular to sample edge.---} Again we refer to Figs.\
\ref{fig:junction_mirror_zigzag} and \ref{fig:junction_mirror_armchair} for a
schematic picture. 
  The presence of an electric field at a $pn$ junction breaks both the sublattice antisymmetry and the mirror symmetry, but the product ${\cal C M T}$ of sublattice conjugation, mirror reflection, and time-reversal remains a good antisymmetry even in the presence of a magnetic field. This combined symmetry operation exchanges scattering states incident from/going to the $p$ and $n$ parts of the junction and changes the valley isospinor as
  \begin{equation}
    \label{eq:isospin_mirror}
    \vnu \to \bar \vnu = \left\{ \begin{array}{ll} {\cal M}_x \vnu & \mbox{zigzag}, \\
    {\cal I} \vnu & \mbox{armchair}, \end{array} \right.
  %
  \end{equation}
  where ${\cal M}_x$ and ${\cal I}$ are mirror reflection in the $x=0$ plane and inversion on the Bloch sphere, respectively, and where the origin $O$ is chosen symmetrically with respect to the mirror axis. Note that these symmetry operations imply that edge states coming in from/going out to the $p$ and $n$ parts of the junction have the same valley isospin for a $pn$ junction with zigzag edges, since $\vnu$ is along the $z$ axis in that case, but oppositely oriented valley isospins for a junction with armchair edges. Since the valley isospins of the two interface states must be oppositely oriented, ${\cal CMT}$ antisymmetry implies that $\vnu$ must be along the $x$ axis for the interface state in a zigzag $pn$ junction.

  {\em Inversion center.---} For a junction with inversion symmetry ${\cal I}$ the presence of an electric field at the $pn$ interface and a magnetic field preserves the antisymmetry under ${\cal C I}$. The corresponding valley isospin change is 
  \begin{equation}
     \vnu \to {\cal R}_{y,\pi} \vnu
  \end{equation}
  for both zigzag and armchair termination, where ${\cal R}_{y,\pi}$ is a $\pi$ rotation around the $y$ axis in the Bloch sphere. As before, the valley isospin after inversion is defined with respect to the inversion image of the origin.

\begin{figure}
    \includegraphics[width=0.9\columnwidth]{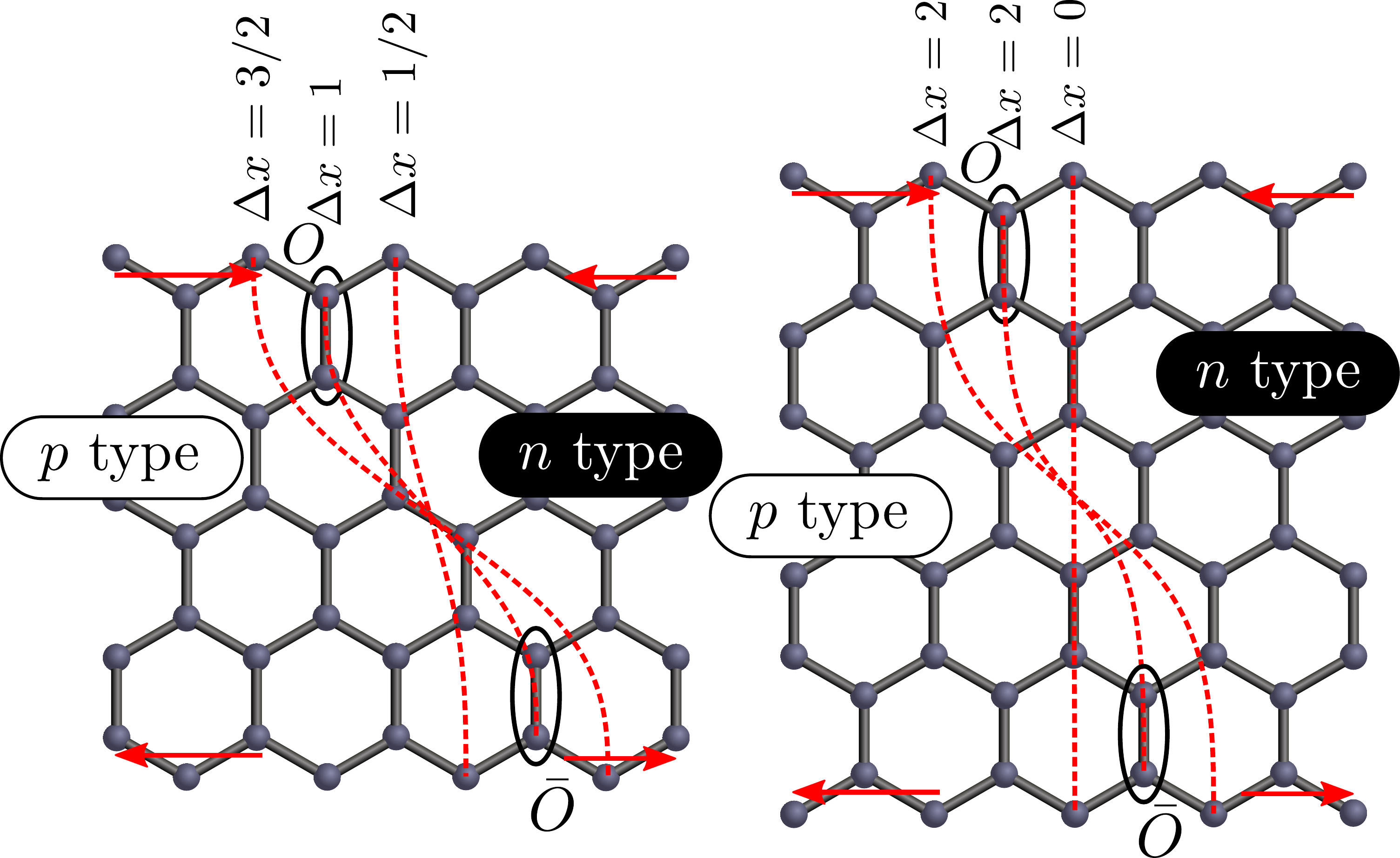}
    \caption{Zigzag $pn$ junctions with an inversion center. The integer $n$ counts the number of half hexagons between the intersection of the $pn$ interface (dashed) and the left and right sample boundaries. Origins $O$ and $\bar O$ compatible with the boundary conditions at the left and right sample boundaries and symmetrically placed with respect to the $pn$ interface are shown for $\delta x=3/2$ (left) and $\Delta x=1$ (right). }
  \label{fig:junction_inversion}
  \end{figure}

  {\em Conductance.---} The conductance of high-symmetry $pn$ junctions with zigzag edges can be determined from symmetry considerations, provided the model has sublattice antisymmetry. These symmetry arguments can be applied to a junction with two mirror axes, as shown in Fig.\ \ref{fig:junction_mirror_zigzag}, and for a junction with an inversion center, if the $pn$ interface is perpendicular to the sample edges and meets the sample boundary at a high-symmetry point, shown schematically in Fig.\ \ref{fig:junction_inversion}. For the case of a junction with zigzag edges and two mirror axes, $\vnu_{\rm in} = \vnu_{\rm out}$, so that the junction is fully reflecting, $G=0$.\cite{tworzydlo2007} For the case of an inversion-symmetric junction (with the additional conditions listed above), the symmetry considerations discussed above fix the valley isospin of an interface state evolving from the chiral edge state coming in from the $p$ region to be $\pm \ve_x$. Inversion symmetry then determines that the interface state evolving into the chiral edge state going into the $n$ region is $\pm \ve_x$. The conductance of the system then depends on the translation vector $\vr_{O\bar O}$ between origins at the two edges. Since a translation by half a hexagon gives a rotation around the $z$ axis by an angle $2 \pi/3$, it follows that the conductance $G$ of an inversion-symmetric zigzag ribbon of a width of $n_{\rm hex}$ hexagons is
\begin{equation}
  G = \frac{e^2}{h} \times \left\{ \begin{array}{ll}
  [1 + \cos(2 \pi \Delta x/3)], & \mbox{if $n_{\rm hex}$ even}, \\
  {}[1 - \cos(2 \pi \Delta x/3)], & \mbox{if $n_{\rm hex}$ odd},
  \end{array} \right. \label{eq:Ginversion}
\end{equation}
where $\Delta x$ is the distance between the intersection points of the $pn$ junction at the opposing zigzag edges, see Fig.\ \ref{fig:junction_inversion}. [Note that Eq.\ (\ref{eq:Ginversion}) is derived for integer and half-integer $\Delta x$ only.]

  The combined ${\cal CMT}$ antisymmetry for a $pn$ junction (with mirror line parallel to the $pn$ interface) also implies a symmetry constraint for the scattering matrix $S$ of the junction as a whole. The constraints on $S$ follow upon noting that the combined ${\cal CMT}$ operation interchanges the incoming scattering states, but does not mix incoming and outgoing scattering states. For a junction with zigzag edges one has $({\cal CMT})^2 = 1$, which leads to the constraint
  \begin{equation}
    S = \begin{pmatrix} 0 & 1 \\ 1 & 0 \end{pmatrix}
    S^* \begin{pmatrix} 0 & 1 \\ 1 & 0 \end{pmatrix}.
    \label{eq:Scondition}
  \end{equation}
This is the same symmetry condition as the one found for Andreev reflection
from superconductors with broken spin degeneracy (Altland-Zirnbauer symmetry
class D\cite{altland1997}).\cite{beri2009} The only $2 \times 2$ unitary
matrices compatible with the condition (\ref{eq:Scondition}) describe either a
fully reflecting junction ($G = 0$) or a fully transmitting junction ($G = 2
e^2/h$).\cite{beri2009} Since a junction cannot be simultaneously fully
reflecting and fully transmitting, it follows that Eq.\ (\ref{eq:Scondition})
imposes a {\em topological} constraint on the junction conductance $G$
evaluated precisely at half filling (energy $\varepsilon = 0$): $G$ does not
change upon continuously deforming the Hamiltonian, as long as the ${\cal CMT}$
antisymmetry is preserved. In particular for a $pn$ junction
with zigzag edges and ${\cal CMT}$ symmetry it follows
that the conductance at the first
quantized Hall plateau is the same as the conductance in the absence of a
magnetic field. The case without magnetic field was considered previously by
Akhmerov {\em et al.},\cite{akhmerov2008} who found that $G=2 e^2/h$ if the
width of the graphene nanoribbon corresponds to an odd number of hexagons and
$G=0$ if the width corresponds to an even number of hexagons. The former
observation is consistent with the results obtained from the symmetry analysis,
see the preceding discussion. The latter observation gives information that
goes beyond what can be obtained from the symmetry analysis alone: It means
that the valley isospin of the interface state evolving from a chiral edge
state coming in from the $p$ region changes from $\pm \ve_x$ to $\mp \ve_x$
upon shifting the intersection of the $pn$ interface and the zigzag edge
between high-symmetry points, as indicated schematically in Fig.\ \ref{fig:junction_mirror_zigzag}. In particular, it follows
that a junction with mirror axis parallel to the $pn$ interface has conductance $G = 2 e^2/h$ if the width of the junction is an even
number of hexagons.\cite{tworzydlo2007}

\section{Numerical calculation of the valley isospin}\label{sec:numerics}

\begin{figure}
  \includegraphics[width=0.8\columnwidth]{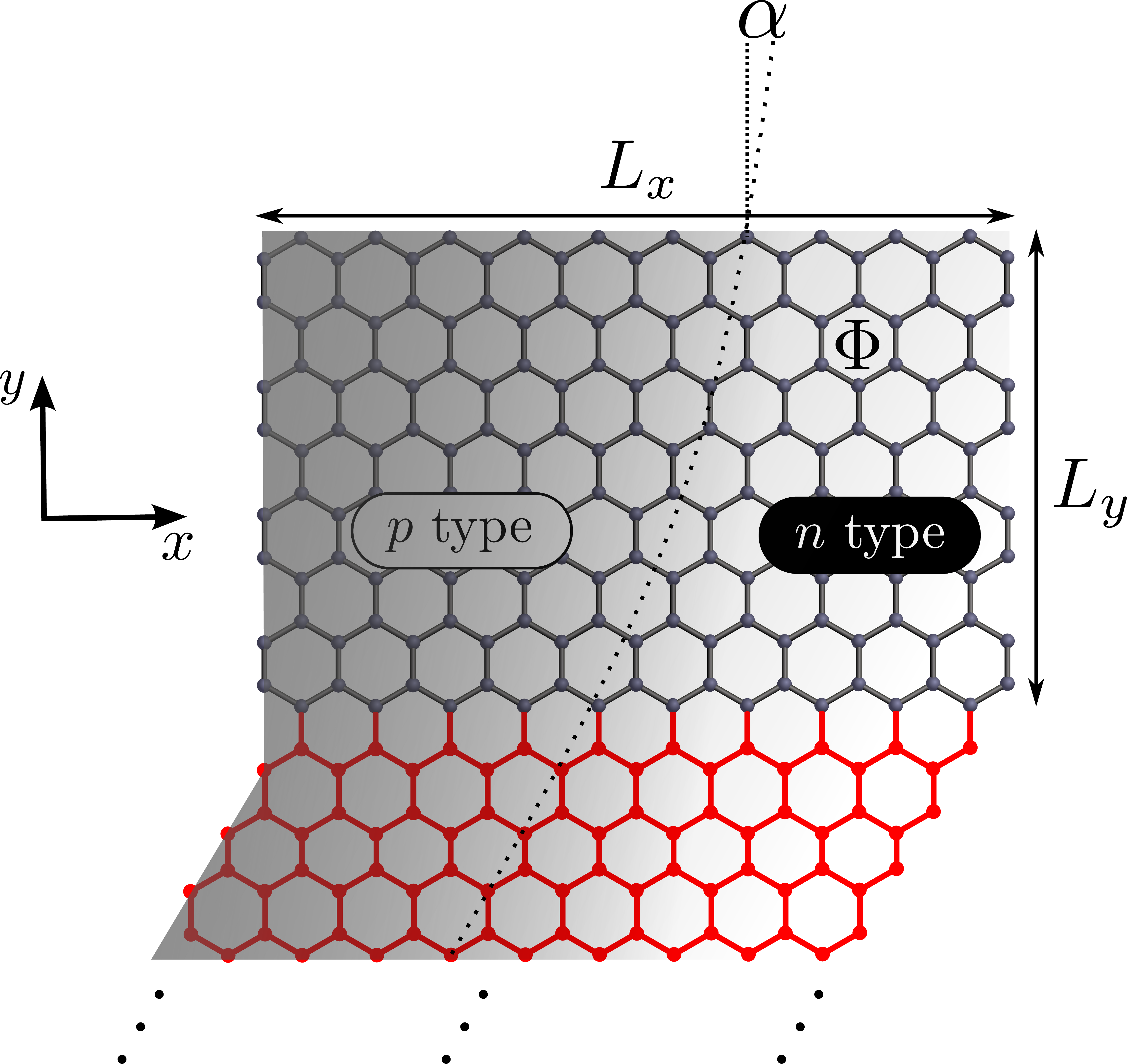}
  \caption{Scattering geometry used to numerically calculate the valley isospin $\vnu$ of the interface states. The lead region is shown in red. The grey background schematically indicates the height of the potential $U(\vr)$, which determines the position of the $pn$ interface. In the ideal lead, the $pn$ interface is parallel to the lead's zigzag boundary; in the scattering region, the $pn$ interface is slightly curved, such that it intersects the zigzag boundary of the scattering region at an angle $\alpha$ to the boundary normal.}
\label{fig:setup}
\end{figure}

To numerically calculate the valley isospin of the interface states we consider the ``stub geometry'' shown in Fig.\ \ref{fig:setup}. A single ideal graphene lead with zigzag edge termination is coupled to a scattering region, to which no other leads are attached. A potential $U(\vr)$ divides the lead and the scattering region into a $p$-type and $n$-type region in such a way that the $pn$ interface runs approximately through the center of the lead and the scattering region.
In the ideal lead, the $pn$ interface is parallel to the lead boundaries. The orientation of the $pn$ interface is gradually changed in the scattering region, so that it makes an angle $\alpha$ with the surface normal at the point where it intersects the zigzag boundary of the scattering region, see Fig.\ \ref{fig:setup}.
A uniform magnetic field is applied, such that the system is in the first quantized Hall plateau throughout.

The ideal lead hosts two ``incoming'' chiral modes propagating along the left and right edge as well as an ``outgoing'' valley-degenerate mode propagating along the $pn$ interface. The scattering matrix of the entire system is calculated using the kwant software package.\cite{groth2014} The valley isospin $\vnu = \vnu_{\rm in}$ of the interface state originating from the edge state coming in on the $p$ side of the $pn$ interface can be immediately read off from the scattering matrix. The valley isospin of the interface state originating from the edge state coming in on the $n$ side of the junction is $-\vnu$. 

The advantage of this geometry over the previously considered slab geometry~\cite{tworzydlo2007,akhmerov2008,handschin2017} is that we can directly access the scattering matrix between the two (incoming) edge states and the two (outgoing) interface states, as described above. The orientation of the lead along zigzag edge is chosen for technical convenience, as it allows for an easy determination of the valley isospin of the two interface states using the longitudinal momentum in the ideal lead. 

The system is described by the tight-binding Hamiltonian
\begin{align}
  H &= \sum_{\vr,\ \vr'\ {\rm n.n.}} t \vert \vr \rangle \langle \vr' \vert +
  \sum_{\vr} U(\vr) \vert \vr \rangle \langle \vr \vert,
  \label{eq:H}
\end{align}
where the sums are over the sites $\vr$ of the hexagonal lattice. The potential $U(\vr)$ defines the $pn$ interface,
\begin{align}
  \label{eq:Ur}
  U(\vr) =&\, \frac{U_0}{4} \left[ f(\vn_{\alpha} \cdot \vr/l_{\rm pn}) (1 + f(y/\xi))
  \right. \nonumber \\ &\, \left. \mbox{}
  + f(\vn_{\pi/6} \cdot \vr/l_{\rm pn}) (1-f(y/\xi)) \right],
\end{align}
where $f(x)$ is a function that smoothly interpolates between $-1$ for $x \ll -1$ and $1$ for $x \gg 1$ and $\vn_{\alpha} = (\cos \alpha,-\sin \alpha)$. [In practice, one may take $f(x) = \tanh x$.] We refer to Fig.\ \ref{fig:setup} for the definition of the $x$ and $y$ directions. Throughout we take the width of a hexagon as the unit of length. The potential (\ref{eq:Ur}) interpolates smoothly between a $pn$ interface parallel to the lead's zigzag edge and an interface intersecting the zigzag edge of the scattering region at angle $\alpha$ with the boundary normal. The lengths $l_{\rm pn}$ and $\xi$ determine the width of the $pn$ junction and the length scale over which the orientation of the $pn$ interfaces between the angles $\pi/6$ (in the lead) and $\alpha$ (at the intersection with the sample boundary). Both lengths must be much larger than the lattice spacing for the continuum description to hold. We further require that $\xi \ll L_y$, where $L_y$ is the depth of the scattering region, see Fig.\ \ref{fig:setup}. For convenience, we have set $l_{\rm pn} = \xi$ in our numerical simulations, but we made sure that our results are independent of this choice as long as the above conditions on $l_{\rm pn}$ and $\xi$ are met. The constant magnetic field is included in the tight-binding description by the Peierls substitution to the hopping amplitudes, such that the magnetic flux through each hexagon is $\Phi$. The value of $\Phi$ is chosen such that the magnetic length $l_{\rm } \sim 1/\sqrt{\Phi} \gg 1$ is much larger than the lattice spacing, whereas the magnetic unit cell is much smaller than the lead width, $L_x \Phi\gg 2\pi$. In practice the latter condition forces us to work with scattering regions of rectangular shape, $L_x \gg L_y$. Finally, we include breaking of sublattice antisymmetry by the inclusion of an additional potential $\delta U$ on the outermost boundary sites. 

\begin{figure}
  \includegraphics[width=\columnwidth]{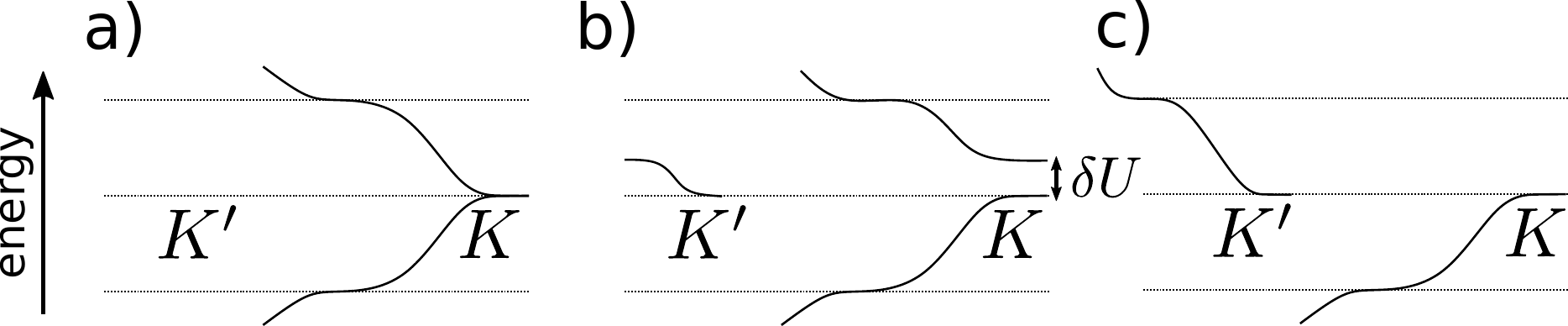}
\caption{Dispersion relation of the quantum Hall edge states (solid curves) along a zigzag edge in the lowest quantized Hall plateau for an edge without sublattice antisymmetry (a), for an edge with sublattice antisymmetry-breaking perturbation $0 < \delta U < E_1$ (b), and for an edge with $\delta U > E_1$ (c), where $E_1$ is the energy of the first Landau level. The dashed lines give the energies of the lowest Landau levels. The chiral edge states in $p$ and $n$ type regions have equal valley isospin in case (a), but opposite isospin in cases (b) and (c). } \label{fig:edge_dis}
\end{figure}

\textit{Zigzag edge: Isospin-conserving vs.\ non-conserving regime.---}
As anticipated in the introduction, the scattering from chiral edge states into interface states depends on the presence or absence of sublattice antisymmetry. In the
presence of sublattice antisymmetry, both incoming edge states have the same
valley isospin, see Eq.~(\ref{eq:zigzagboundary}) and 
Fig.~\ref{fig:edge_dis}a. Correspondingly, valley isospin is not conserved
when the chiral edge states are converted into valley-degenerate interface
states at the intersection of the $pn$ interface and the sample edge. 
The non-conservation of the valley isospin at a smooth interface is 
associated with the fact that at zero energy ({\em i.e.}, precisely at
the $pn$ interface) the edge state has longitudinal momentum $k=\pi$ and 
is localized on the outermost layer of lattice sites only.~\cite{akhmerov2008} 
After introducing a perturbation that breaks the sublattice antisymmetry, 
such as next-nearest-neighbor hopping or a local 
onsite potential along the edge, the edge state acquires a finite energy 
$\delta U$ at $k = \pi$ and the dispersion relation of the edge states is 
changed. Typical dispersions for the case of broken chiral antisymmetry 
are shown in Figs.~\ref{fig:edge_dis}b and c. Most importantly, for broken 
sublattice antisymmetry the two incoming edge states at the same edge
have opposite valley polarization. For a smooth $pn$ interface valley isospin 
is conserved in the transition from chiral edge states to interface states.

\begin{figure}
  \includegraphics[width=\columnwidth]{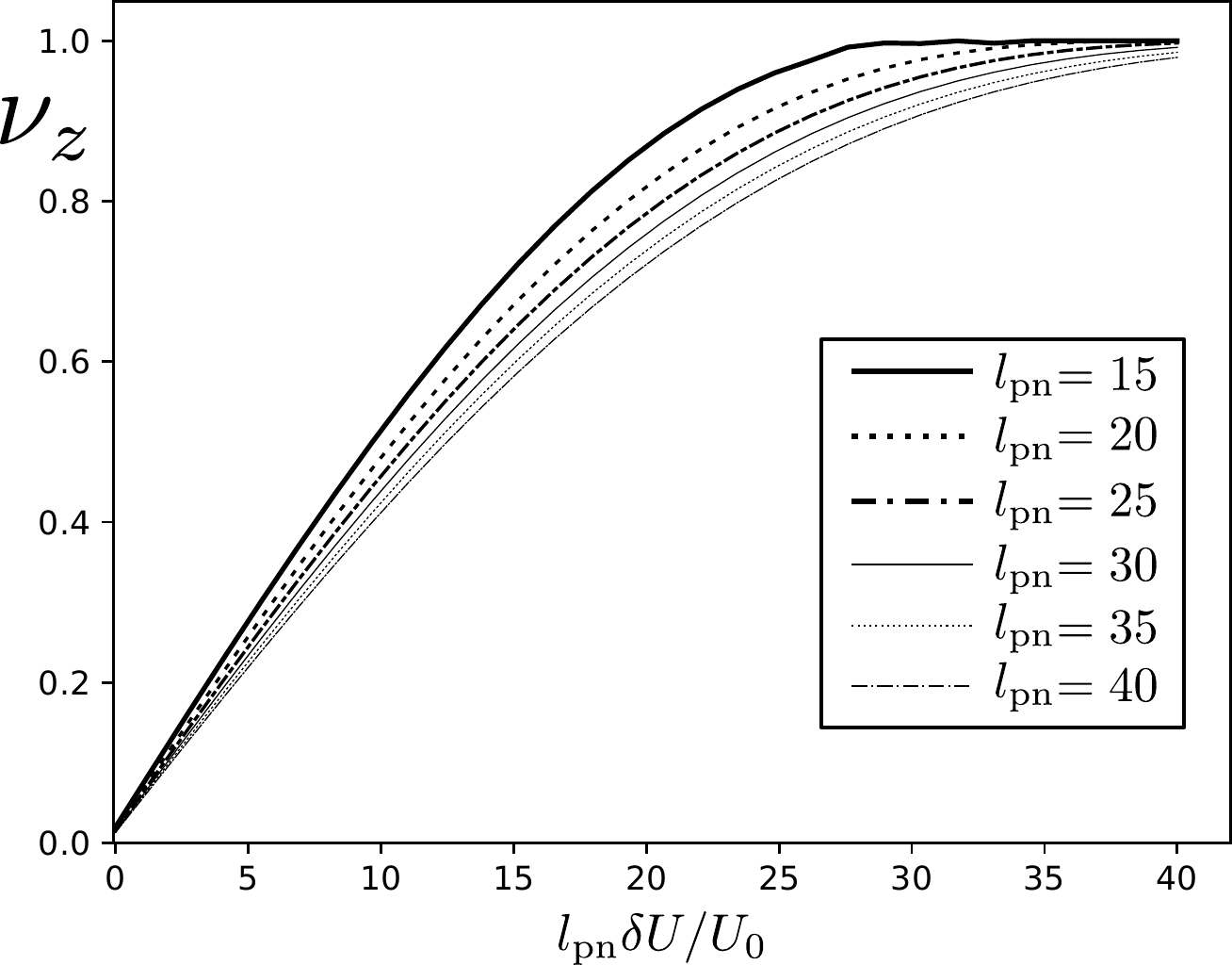}
  \caption{The $z$-component $\nu_z$ of the isospin of interface states
    originating from the $p$-type region, as a function of $l_{\rm pn} \delta
    U/U_0$. We have chosen $U_0 = 0.2 t$ and the values of $l_{\rm pn}$
  (measured in units of the hexagon width) are given in the inset. The value of
the magnetic field is such that the first Landau level is at $E_1 = 0.2t$ and
the size of the scattering region $L_x=360$ and $L_y=10l_{\rm pn}$. The
sublattice antisymmetry is broken by onsite potential of magnitude $\delta U$
that is added locally on the outermost
sites of the zigzag edge. }

\label{fig:deltaU}
\end{figure}

The transition between the two regimes is illustrated in Fig.~\ref{fig:deltaU}, which shows the $z$ component of the valley isospin of the interface state $\vnu$. In the presence of sublattice antisymmetry one has $\nu_z = 0$, consistent with the symmetry arguments of the previous Section; if sublattice antisymmetry is broken, which requires that the sublattice antisymmetry-breaking energy scale $\delta U$ be sufficiently large in comparison to the potential difference $U_0/l_{\rm pn}$ on neighboring lattice sites, $\nu_z$ approaches the unit valley isospin of the incoming edge state. 

To estimate the energy scale $\delta U$ for graphene we note that $\delta U = t'$ in a lattice model with next-nearest neighbor hopping amplitude $t'$. The experimental estimate $t' \approx 0.3$eV,\cite{kretinin2013} is larger than the distance to the first Landau level $E_1$ at experimentally relevant magnetic field, $E_1 \sim 0.1$ eV for magnetic fields $\sim 5$ T. Since the total potential drop across the $pn$ junction $U_0$ must be below $E_1$ for a sample at the lowest quantized Hall plateau, it follows that $\delta U \gg U_0$, so that the condition $\delta U \gg U_0/l_{\rm pm}$ is met even for a relatively sharp $pn$ interface. In practice, to reach the isospin non-conserving regime would require the addition and fine tuning of an additional edge potential, to offset the energy shift $\delta U$ from next-nearest-neighbor hopping. The isospin non-conserving regime, despite the experimental difficulty it poses, contains interesting physics. In particular it allows for a complete control of the valley isospin of the interface states by purely electrical means, as we now discuss.

\begin{figure}
  \includegraphics[width=\columnwidth]{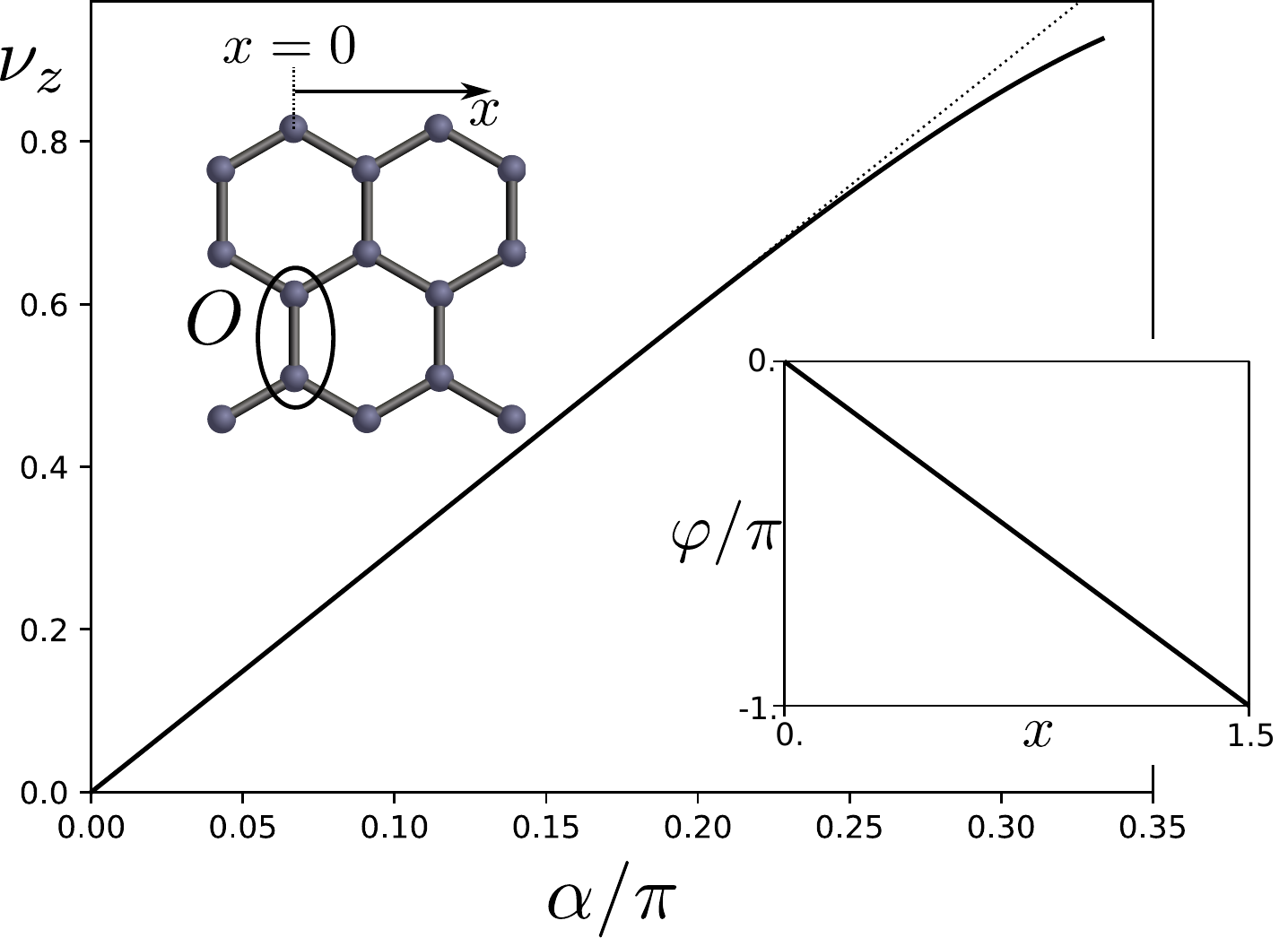}
  \caption{Main panel: Valley isospin component $\nu_z$ of the interface state originating from the $p$-type region as a function of the angle $\alpha$ between the $pn$ interface and the boundary normal. The dashed line shows the linear approximation (\ref{eq:Zalpha}). Right inset: The azimuthal angle $\varphi$ of the valley isospin $\vnu$ at $\alpha = 0$, as a function of the $x$ coordinate of the intersection of the $pn$ junction and the zigzag sample boundary. The coordinate $x$ is measured in units of the hexagon width; the origin $x=0$ and the reference unit cell are chosen at one of the outermost lattice sites, as shown in the left inset. } \label{fig:azimuthal}
\label{fig:alpha}
\end{figure}

\textit{Zigzag edge: isospin non-conserving regime.---} We first focus on the
case $\alpha = 0$, the $pn$ interface being perpendicular to the zigzag edge of
the scattering region. For mirror-symmetric positions of the $pn$ interface, we
expect that $\vnu$ is in the $xy$ plane, see the previous Section. Our
numerical calculations confirm that this continues to be the case for arbitrary
position of the $pn$ interface (data not shown). Numerical results for the
azimuthal angle of $\vnu$ in the $xy$ plane are shown in Fig.\
\ref{fig:azimuthal}. Within numerical accuracy, these results can be described
by the simple equation
\begin{align}
  \vnu= \ve_x \cos (2 x \pi/3)- \ve_y \sin(2 x \pi/3),
  \label{eq:azimuthal}
\end{align}
where $x$ is the position of the $pn$ interface in the units of the hexagon width, with $x=0$ placed at one of the outermost sites on the edge and the valley isospin is defined with respect to a unit cell at $x=0$ (see Fig.\ \ref{fig:azimuthal}, inset). Note that Eq.\ (\ref{eq:azimuthal}) does not have the periodicity of the hexagonal lattice, because the isospin in Eq.\ (\ref{eq:azimuthal}) is defined with respect to a fixed choice of the reference unit cell. To restore the periodicity of the lattice, we must calculate the valley isospin with respect to a reference unit cell that moves along with the $pn$ interface. We recall that increasing $x$ by one corresponds to a rotation by $2 \pi/3$ around the $z$ axis, see Eq.\ (\ref{eq:translation}), so that the valley isospin is indeed periodic when calculated with respect to a reference unit cell that shifts simultaneously with the $pn$ interface. We also note that shifting the position reference unit cell by half a hexagon corresponds to a rotation of the valley isospin by $-2 \pi/3$ around the $z$ axis, see Eq.\ (\ref{eq:translation}), so that the valley isospin changes sign when calculated with respect to a simultaneously shifted unit cell for a translation by half a hexagon, consistent with the symmetry considerations of Sec.\ \ref{sec:model}.

Upon varying the angle $\alpha$, the azimuthal angle remains a fast function of the precise location of the intersection of the $pn$ interface and the sample edge. At the same time, $\vnu$ acquires a nonzero $z$ component $\nu_z$, see Fig.\ \ref{fig:alpha}. For an interface slanted towards the $p$ region (positive $\alpha$), $\nu_z$ is positive, which can understood from the observation that for sharp angles the edge state and the interface state ``gap out'', so that electrons coming from the edge no longer reach the junction itself, which is where isospin violation takes place, but deflect before that, thus preserving part of their valley identity. For $\alpha \lesssim \pi/4$, $\nu_z$ in Fig.\ \ref{fig:alpha} is an approximately linear function of $\alpha$,
\begin{align}
  \label{eq:Zalpha}
  \nu_z \approx c \alpha,
\end{align}
where $c \approx 0.96$. 

\begin{figure}
  \includegraphics[width=0.8\columnwidth]{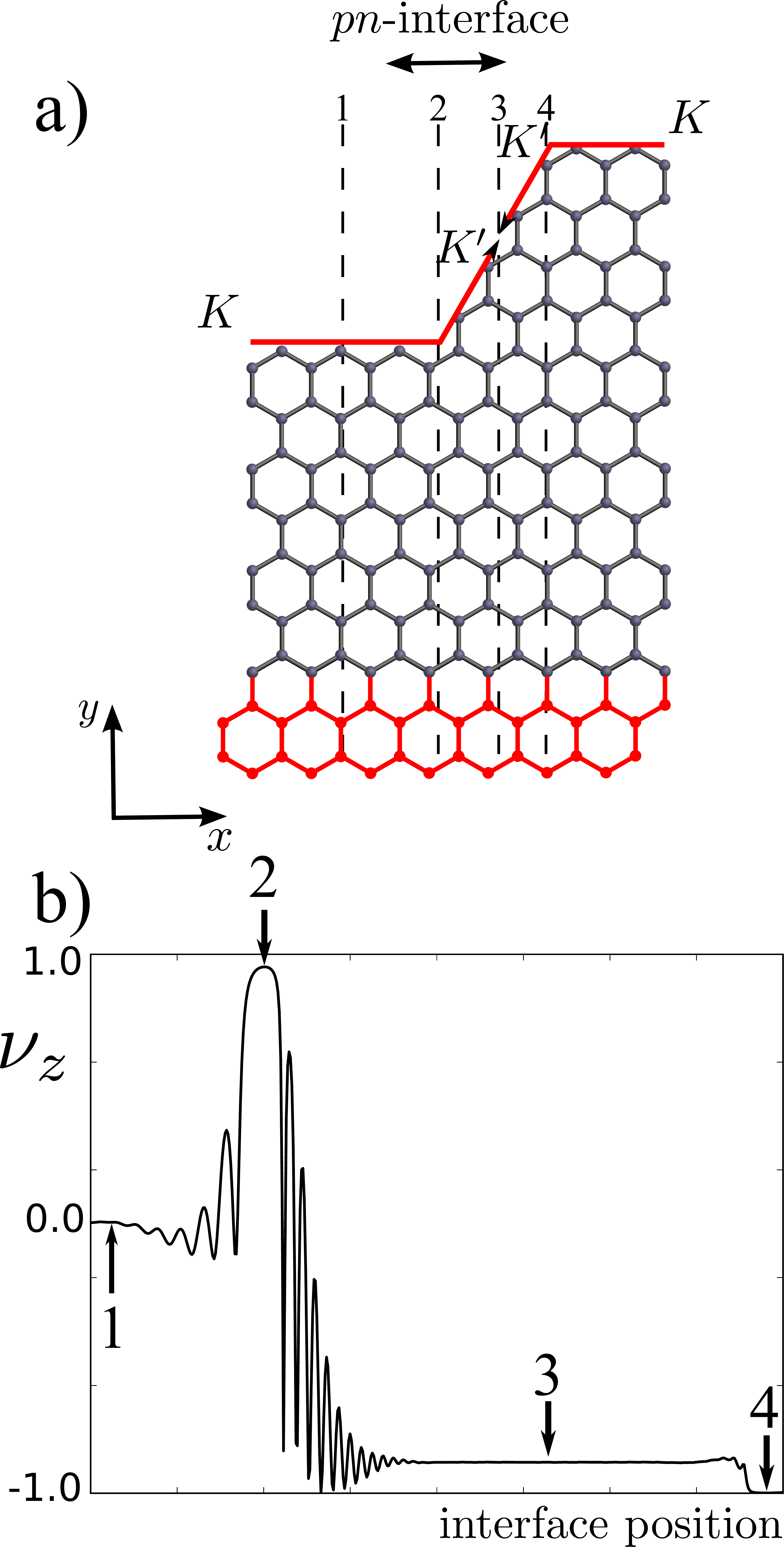}
  \caption{(a) Scattering region with two different zigzag terminations at the upper edge. The valley polarization of the chiral edge states is $\ve_z$ ($K$ valley) for the ``horizontal'' edges and $-\ve_z$ ($K'$ valley) for the ``tilted'' edge, as indicated in the figure. (b) Valley isospin $\nu_z$ of the interface state originating from the (left) $p$-type region. The numbers ``1'', ``2'', ``3'', and ``4'' indicate positions of the $pn$ interface shown in panel (a). }
\label{fig:setup1}
\end{figure}

An interesting geometry in which the $\alpha$ dependence of $\nu_z$ is illustrated is shown in Fig.\ \ref{fig:setup1}a. Here, the upper edge of the scattering region alternates between a ``horizontal'' zigzag edge with edge states in the $K$ valley ($\vnu = \ve_z$) a ``tilted'' zigzag edge with chiral edge states in the $K'$ valley ($\vnu = -\ve_z$). Figure \ref{fig:setup1}b shows the result of a numerical calculation of $\nu_z$ as a function of the position of the $pn$ interface, where the $pn$ interface remains ``vertical'' throughout (as indicated by the dashed lines in Fig.\ \ref{fig:setup1}a). When the interface is at the position ``$1$'', we find $\nu_z = 0$, as discussed above. At position ``2'' the two incoming edge states have opposite values of isospin, thus it is possible to conserve the valley isospin, giving rise to $\nu_z = 1$, despite the non-smoothness of the lattice boundary at this point. At position ``3'' the angle with the surface normal is $\alpha = -\pi/3$, giving a negative value $\nu_z \approx -0.90$ (compare with Fig.~\ref{fig:alpha}). Finally, when the interface reaches position ``4'', we again expect that valley isospin is conserved, yielding $\nu_z = -1$ for an interface state originating from the $p$-type region.

\section{Conductance in ribbon with mixed armchair and zigzag edges}\label{sec:extensions}
\begin{figure}
  \includegraphics[width=\columnwidth]{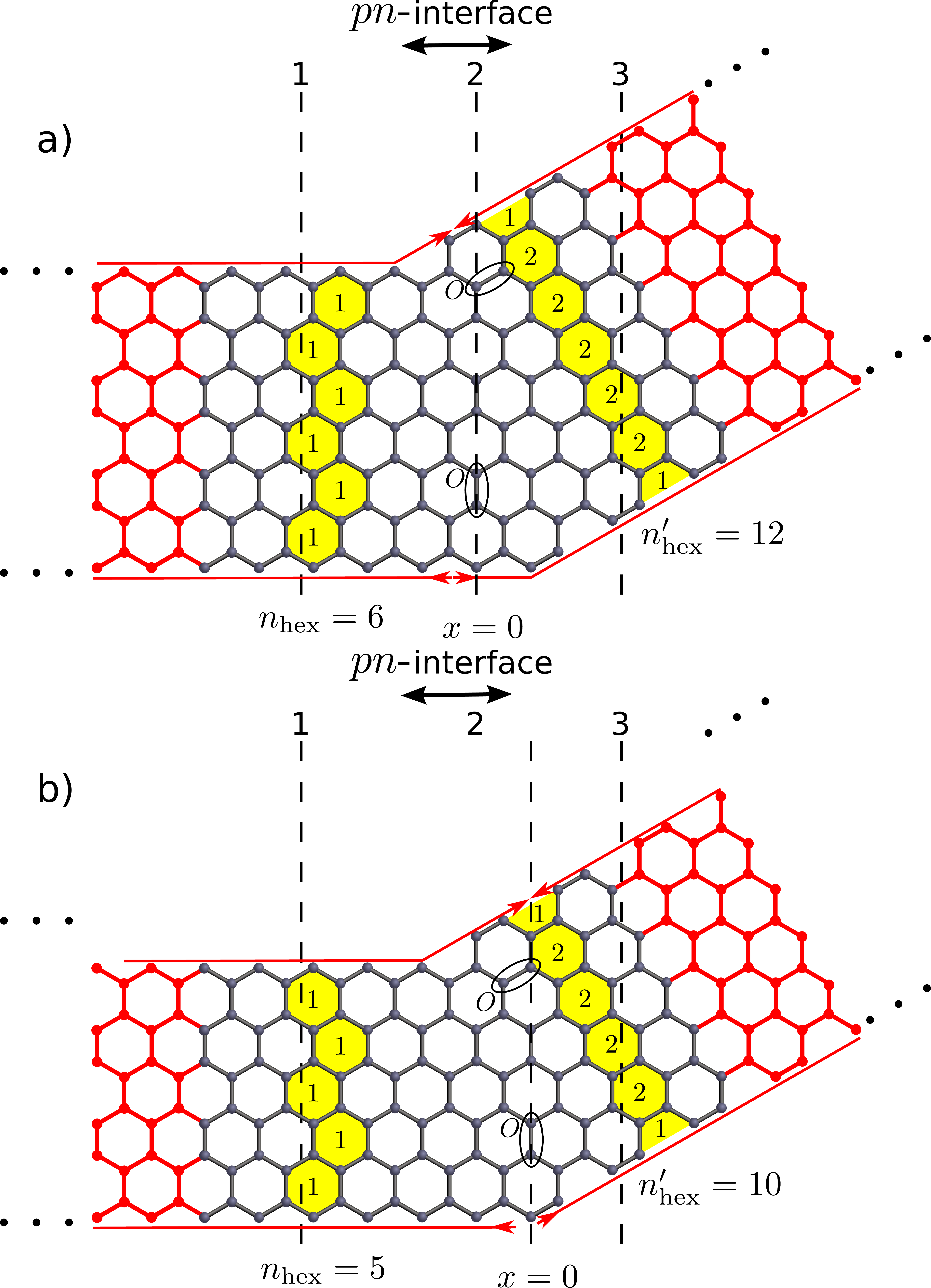}
  \caption{A nanoribbon with sections having zigzag/zigzag (1), zigzag/armchair (2) and armchair/armchair termination (3). With broken sublattice symmetry, the conductance $G$ is different for the three combinations of boundary termination, but does not depend on the precise location or orientation of the $pn$ interface within these three sections. In contrast, in the presence of sublattice symmetry, $G$ depends strongly on the precise orientation (in region 1) or position (in region 2) of the $pn$ interface. The left panel shows a lattice in which the numbers $n_{\rm hex}$ of hexagons in section 1 is even and the number $n_{\rm hex}'$ of half-hexagons in section 3 is a multiple of 3 plus one. The right panel shows a lattice in $n_{\rm hex}$ and $n_{\rm hex}'$ are odd and not a multiple of 3 plus one, respectively. }
\label{fig:zigzag_armchair}
\end{figure}

In this section we apply the theory of the two previous Sections to the
calculation of the conductance of a graphene nanoribbon for three different
combinations of edge terminations, see Fig.~\ref{fig:zigzag_armchair}. We
compare the isospin-conserving regime (broken sublattice antisymmetry) and the
non-conserving regime (unbroken sublattice antisymmetry). The conductance is
calculated using Eq.\ (\ref{eq:conductance}), where we use the results of
Secs.\ \ref{sec:model} and \ref{sec:numerics} for the valley isospin $\vnu_{\rm
in}$ and $\vnu_{\rm out}$ of interface states originating from/evolving into
chiral edge states at the boundary of the $p$-type region. Although the results
of Sec.\ \ref{sec:numerics} were formulated for the isospin $\vnu = \vnu_{\rm
in}$ of an interface state with incoming boundary conditions only, the isospin
$\vnu_{\rm out}$ of an interface state with outgoing boundary conditions can be
obtained using the symmetry relations (\ref{eq:mirror_parallel}).

In the physically relevant regime of broken sublattice antisymmetry, we find
that the conductance for the position ``1'' of the $pn$-interface in
Fig.~\ref{fig:zigzag_armchair} (two zigzag boundaries) is $2 e^2/h$. This
follows from Eq.\ (\ref{eq:conductance}) upon noting that $\vnu_{\rm in}$ and
$\vnu_{\rm out}$ point along the $z$ axis and that $\vnu_{\rm out} = {\cal M}_z
\vnu_{\rm in}$ is the mirror image of $\vnu_{\rm in}$ under reflection in the
$xy$ plane, so that $\vnu_{\rm out} = - \vnu_{\rm in}$. This result holds
independent of the width of the nanoribbon or the orientation of location of
the $pn$ interface. Similarly, for position ``2'' (mixed zigzag and armchair
boundaries) the conductance $G = e^2/h$. This again follows from Eq.\
(\ref{eq:conductance}), using that $\vnu_{\rm in}$ is in the $xy$ plane
(armchair edge), whereas $\vnu_{\rm out}$ points along the $z$ axis (zigzag
edge), so that $\vnu_{\rm in} \cdot \vnu_{\rm out} = 0$. Again, this result is
independent of the width of the nanoribbon or the orientation of location of
the $pn$ interface. Finally, for position ``3'' (armchair edges) the
conductance depends on the ribbon width $n_{\rm hex}'$ measured in
half-hexagons. One has $G = 2 e^2/h$ if $n_{\rm hex}'$ is a multiple of three
plus one and $G = e^2/2h$ otherwise.\cite{tworzydlo2007} Summarizing, we find
\begin{equation}
  G = \frac{e^2}{2h} \times \left\{ \begin{array}{ll} 
  4 & \mbox{for two zigzag edges}, \\
  2 & \mbox {for one zigzag, one armchair edge}, \\
  4\ \mbox{or}\ 1 & \mbox{for two armchair edges}.
\end{array} \right.
  \label{eq:Gzigzag_armchair}
\end{equation}

\begin{figure}
  \includegraphics[width=\columnwidth]{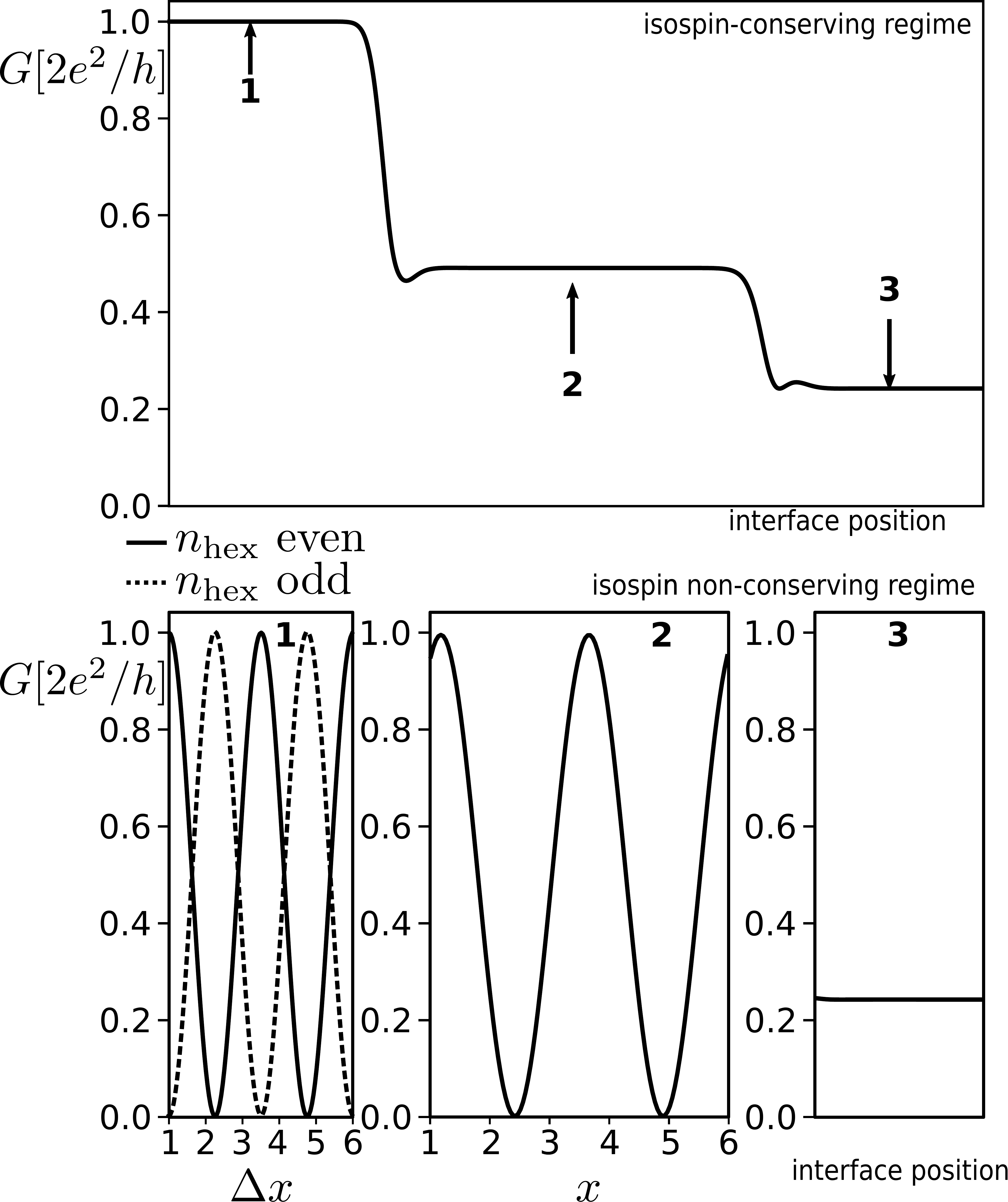}
  \caption{Conductance $G$ if a nanoribbon with mixed boundary conditions, as shown schematically in the left panel (solid) and right panel (dashed) of Fig.\ \ref{fig:zigzag_armchair}. The top and bottom panels are for broken and unbroken sublattice antisymmetry, respectively. The labels ``1'', ``2'', and ``3'' refer to the three regions shown in Fig.\ \ref{fig:zigzag_armchair}: A ribbon with two zigzag edges, a ribbon with one zigzag and one armchair edge, and a ribbon with two armchair edges, respectively. The inset shows how $G$ depends on the orientation of the $pn$ interface for a ribbon with two zigzag edges (region ``1'' in Fig.\ \ref{fig:zigzag_armchair}) in the case of unbroken sublattice antisymmetry.}
\label{fig:Gzigzag_armchair}
\end{figure}

In the opposite regime in which sublattice antisymmetry is present, the
conductance results in positions ``1'' and ``2'' (two zigzag edges and one
zigzag, one armchair edge) are markedly different from the regime of broken
sublattice antisymmetry discussed above; for position ``3'' (two armchair
edges) the presence or absence of sublattice antisymmetry plays no role. 

For position ``1'' one finds that $G$ depends on the junction width $n_{\rm
hex}$, measured in hexagons:\cite{tworzydlo2007} $G = 2 e^2/h$ if $n_{\rm hex}$
is even, $G = 0$ if $n_{\rm hex}$ is odd. Whereas this result is independent of
the {\em position} of the $pn$ interface along the ribbon, the conductance
depends strongly on the {\em orientation} of the $pn$ interface. The reason is
the strong dependence of the azimuthal angle of the isospin $\vnu_{\rm in}$ and
$\vnu_{\rm out}$ of the interface states on the precise location of the
intersection of the $pn$ interface and the sample boundary, see Eq.\
(\ref{eq:azimuthal}). For a (longitudinal) distance $\Delta x$ between the
points where the $pn$ interface intersects the ``top'' and ``bottom'' zigzag
edges, one finds for small intersection angles $\alpha$, that
\begin{equation}
  G = \frac{e^2}{h} [1 + \cos(2 \pi \Delta x/3 + \pi n_{\rm hex})].
  \label{eq:Gzigzag}
\end{equation}
Equation (\ref{eq:Gzigzag}) generalizes the result (\ref{eq:Ginversion})
previously derived for junctions with combined inversion symmetry and
sublattice antisymmetry. Precisely the same result was found by Akhmerov {\em
et al.}\ for the conductance of a zigzag graphene $pn$ junction in zero
magnetic field.\cite{akhmerov2008} On a heuristic level, we can understand this
coincidence as resulting from the fact that the origin of the
valley-isospin-nonconserving processes is the same in both cases: The chiral
edge states are localized on the atomic scale precisely at the $pn$ interface
if sublattice antisymmetry is present. For values of $\Delta x$ that are
compatible with an inversion symmetric position of the $pn$ interface, the
agreement between the two results follows from the symmetry considerations of
Sec.\ \ref{sec:model}, see the discussion following Eq.\ (\ref{eq:Scondition}).

For a $pn$ interface at position ``2'' (one zigzag edge, one armchair edge),
the conductance $G$ depends strongly on the position of the $pn$ interface, but
only weakly on its orientation. If the $pn$ interface intersects the zigzag
edge perpendicularly (angle $\alpha = 0$), one has $\vnu_{\rm out} = \ve_x
\cos(2 \pi x/3) - \ve_y \sin(2 \pi x/3)$, see Eq.\ (\ref{eq:azimuthal}), and
$\vnu_{\rm in}= {\cal R}_{2 \pi/3}(-\ve_y) = \ve_x \cos(\pi/6) + \ve_y
\sin(\pi/6)$, where $x$ is the position of the intersection of the $pn$
interface and the zigzag edge, measured with respect to the reference position
$x=0$, oriented perpendicular to the zigzag edge, see
Fig.~\ref{fig:zigzag_armchair}. This gives
\begin{equation}
  G = \frac{e^2}{h} [1 + \cos(\pi/6-2 \pi x/3)].
  \label{eq:prediction2}
\end{equation}
The fast oscillations as a function of the position $x$ of the $pn$ interface
persist if the $pn$ interface is not orthogonal to the zigzag edge
(angle $\alpha \neq 0$), although the amplitude of the oscillations decreases
because the valley isospin of the interface state associated with the zigzag
edge acquires a finite $z$ component, see Fig.\ \ref{fig:azimuthal}. 

In Fig.\ \ref{fig:Gzigzag_armchair} we compare these theoretical predictions
with the result of a numerical calculation of the conductance of the nanoribbon
using the kwant software package.\cite{groth2014} For position ``1'' (two
zigzag boundaries) the agreement between numerical simulations and the
theoretical predictions is excellent, both in the isospin-conserving and in the
isospin-non-conserving regime. For positions ``2'' and ``3'', which contain at
least one armchair boundary, upon taking the continuum limit (width $l_{\rm
pn}$ of the $pn$ junction and magnetic length $l$ much larger than lattice
constant $a$), we find that the convergence to the theoretical result is much
slower than for a graphene nanoribbon with zigzag boundaries. The agreement
between simulation and theory could be improved upon smoothly taking the
magnetic field to zero at the sample boundaries. This ensures that the
isospin-conserving boundary condition (\ref{eq:armchairboundary}) continues to
be valid in the presence of a magnetic field. Such a smoothly vanishing
magnetic field was used in the simulations shown in Fig.\
\ref{fig:Gzigzag_armchair}, for which the disagreement between simulation and
theory is in the range of a few percent only. Without a smoothly vanishing
magnetic field at the armchair boundaries, no quantitative agreement between
the theoretical predictions (\ref{eq:Gzigzag_armchair}) and
(\ref{eq:prediction2}) and the numerical simulations could be obtained (data
not shown). We note that a significant difference between the theoretical
prediction and numerically observed conductance value was previously seen in
Ref.\ \onlinecite{tworzydlo2007}. Remarkably, in Ref.\
\onlinecite{tworzydlo2007} the agreement between the theoretical prediction
based on isospin conservation and the numerical simulations improve upon going
towards an abrupt $pn$ interface, which is outside the parameter regime in
which one would expect isospin conservation to hold.

\section{Conclusion}
\label{sec:conclusion}

Understanding the valley isospin of chiral interface at a graphene $pn$ junction in a quantizing magnetic field is a key element of a theory of the transport properties of such a junction\cite{tworzydlo2007} and, in a grander scheme, a necessary step towards establishing such junctions as a ``valleytronic'' device.\cite{sekera2017,handschin2017} We have shown that, for a $pn$ interface in a graphene sheet with one or more zigzag edges, the presence or absence of a sublattice antisymmetry strongly affects the valley isospin of interface states. Most theoretical studies in the literature consider simplified lattice models with nearest-neighbor hopping only, which possess a sublattice antisymmetry. The sublattice antisymmetry is not present in realistic models of graphene, however, and experiments show that the energy scale associated with sublattice-antisymmetry breaking is large in comparison to the Landau level spacing or the potential step in a $pn$ junction.\cite{kretinin2013} We therefore expect that --- as far as real devices with zigzag edges are concerned --- the case of (strongly) broken sublattice antisymmetry is relevant for the description of experiments on graphene $pn$ junctions at the first quantized Hall plateau,\cite{williams2007,lohmann2009,ki2009,ki2010,woszczyna2011,williams2011,schmidt2013,matsuo2015b,klimov2015,matsuo2015,kumada2015,handschin2017,makk2018} not theories involving lattice models with nearest-neighbor hopping only.\cite{tworzydlo2007,handschin2017}

{}Nevertheless, from a theoretical point of view, the case of unbroken sublattice antisymmetry is the more interesting one, as it features a strong dependence of the valley isosopin of the interface states on the precise position or orientation of the $pn$ interface. On the one hand, such a dependence on the position of the $pn$ interface offers the possibility to manipulate valley isospin using purely electrostatic means. On the other hand, it also signals an extreme sensitivity of the valley isospin and the conductance of a nanoribbon to microscopic details: The valley isospin $\vnu$ of the interface states rotates by a large angle $\sim \pi$ if the position of the $pn$ interface shifts by only one lattice spacing. The sensitivity to the precise position of the $pn$ interface limits the possibility to {\em a priori} predict the valley isospin $\vnu$, although it still leaves room for an {\em a posteriori} fine tuning of $\vnu$. The strong dependence of $\vnu$ on the position of the $pn$ interface also explains the extreme disorder sensitivity seen in previous numerical simulations of the nearest-neighbor model.\cite{tworzydlo2007,handschin2017}

As argued above, the case of broken sublattice antisymmetry is the physically relevant one. In this regime, the expression (\ref{eq:Gzigzag_armchair}) for the conductance of a graphene $pn$ junction in the first quantized Hall plateau is markedly different from the results in the presence of sublattice symmetry, see Refs.\ \onlinecite{tworzydlo2007,handschin2017}. Moreover, unlike in the case of unbroken sublattice antisymmetry, these results for the conductance are robust to small changes in the position or orientation of the $pn$ interface, so that they should continue to hold in the presence of smooth disorder. The presence of short-range scatterers, which cause intervalley scattering, gives rise to additional isospin rotations of the interface states, see Ref.\ \onlinecite{fraessdorf2016}.

A number of conductance experiments on graphene $pn$ junctions in the first quantized Hall plateau have measured the value $G = e^2/h$, without mesoscopic fluctuations. Although this value of $G$ is consistent with the ensemble average conductance in a strongly disordered junction,\cite{abanin2007,fraessdorf2016} the absence of mesoscopic fluctuations in the experiment is not. We note that the experimental observation of a non-fluctuating conductance $G = e^2/h$ is consistent with our prediction for a ribbon with one zigzag edge and one armchair edge, see Eq.\ (\ref{eq:Gzigzag_armchair}), but also caution that such an explanation is not consistent with shot noise measurements, which find a Fano factor that is significantly below the theoretical expectation for that case.\cite{matsuo2015,kumada2015}

\acknowledgments
We thank Anton Akhmerov for motivating us to study this problem and Peter Silvestrov for stimulating discussions. This work is supported by the German Research Foundation (DFG) in the framework of the Priority Program 1666 ``Topological Insulators''.

\bibliography{refs}
\end{document}